\documentclass[aps,pre,twocolumn,groupedaddress,showkeys]{revtex4}
\usepackage{graphics,color}

\usepackage[]{graphicx}
\DeclareGraphicsExtensions{.pdf}

\newcommand {\e}[1]{\mathrm{~#1}}       
\newcommand {\E}[1]{\cdot 10^{#1}}		
\newcommand {\vek}[1]{\mathbf{#1}}

\begin{document}

\title{Spin glass models for a network of real neurons}

\author{Ga\v{s}per Tka\v{c}ik,\footnote{Present address: Department of Physics and Astronomy, University of Pennsylvania, Philadelphia, Pennsylvania 19104--6396.}$^{a,c}$ Elad Schneidman,\footnote{Present address:  Department of Neurobiology, Weizmann Institute of Science, Rehovot 76100, Israel.}$^{a-c}$ Michael J. Berry II,$^{b}$ and William Bialek$^{a,c,d}$}
\affiliation{$^a$Joseph Henry Laboratories of Physics, $^b$Department of Molecular Biology, 
$^c$Lewis--Sigler Institute for Integrative Genomics,
and $^d$Princeton Center for Theoretical Science\\
Princeton University, Princeton, New Jersey 08544 USA}

\date{\today}

\begin{abstract}
Ising models with pairwise interactions are the least structured, or maximum--entropy, probability distributions that exactly reproduce measured pairwise correlations between spins. Here we use this equivalence to construct Ising models that describe the correlated spiking activity of populations of 40 neurons in the salamander retina responding to natural movies. We show that pairwise interactions between neurons account for observed higher--order correlations, and that for groups of 10 or more neurons pairwise interactions can no longer be regarded as small perturbations in an independent system. We then construct network ensembles that generalize the network instances observed in the experiment, and study their thermodynamic behavior and coding capacity.
Based on this construction, we can also create synthetic networks of 120 neurons, and find that with increasing size the networks operate closer to a critical point and start exhibiting collective behaviors reminiscent of spin glasses. We examine closely two such behaviors that could be relevant for neural code: tuning of the network to the critical point to maximize the ability to encode diverse stimuli, and using the metastable states of the Ising Hamiltonian as neural code words.
\end{abstract}

\maketitle

\section{Introduction}

Physicists have long explored analogies between the statistical mechanics of Ising models and the dynamics of neural networks \cite{hopfield_82,amit_89, hertz_91}.   The goal of this effort has been not to simulate the details of particular networks, but to understand how interesting functions can emerge, collectively, from large populations of neurons.  In the spirit of modern statistical mechanics, one hopes that these collective behaviors will have some degree of universality, and hence that one can make progress without knowing all of the microscopic details of each system.
A classic example of this work is the model of associative or 
content--addressable memory due to Hopfield \cite{hopfield_82}, which is able to recover the correct memory from any of its subparts of sufficient size. Because the computational substrate of neural states in these models were binary `spins,' and the memories were realized as locally stable states of the network dynamics, methods of statistical physics could be brought to bear on theoretically challenging issues such as the storage capacity of the network or its reliability in the presence of noise \cite{amit_89, hertz_91}. 
On the other hand, precisely because of these abstractions, it has not always been clear how to bring the predictions of the models into contact with experiment.


Recently it has been suggested that the analogy between Ising models and neural networks can be turned into a  precise mapping, and connected to experimental data,  using the maximum entropy framework \cite{schneidman+al_06}.  We imagine a neural system exposed to a stationary stimulus ensemble, in which simultaneous recordings of $N$ neurons can be made. In small windows of time,  a single neuron ${\rm i}$ either does ($\sigma_{\rm i}=+1$) or does not ($\sigma_{\rm i} = -1$) generate an action potential or ``spike'' \cite{spikes}; the state of the entire network in that time bin thus is described by a `binary word' or spin configuration 
$\{\sigma_{\rm i}\}$. 
As the system responds to its inputs, it visits each of these states with some probability $P_{\rm expt}(\{\sigma_{\rm i}\})$.  Even before we ask what the different states `mean,' for example as code words in a representation of the sensory world, specifying  this distribution requires us to determine the probability of each of $2^N$ possible states.  In practice, once   $N$ increases beyond $\sim 10$, this becomes impossible. The idea of the maximum entropy construction is to measure low order moments of the distribution, such as the average probability of spiking for each cell ($\langle \sigma_{\rm i}\rangle$) and the correlations between pairs of cells ($C_{\rm ij} = \langle \sigma_{\rm i}\sigma_{\rm j} \rangle  -  \langle \sigma_{\rm i}\rangle\langle\sigma_{\rm j} \rangle$), where the averages are taken over the course of the experiment, and then search for a probability distribution $P(\{\sigma_{\rm i}\})$ that matches these experimental measurements but otherwise is as unstructured as possible.  
Minimizing structure means maximizing entropy \cite{jaynes_57}, and given any set of moments or correlations that we want to match, the form of the maximum entropy distribution is easy to find analytically.  

For the particular case we are interested in, where the states of neurons are given by binary variables $\sigma_{\rm i}$ and we match the mean spike probabilities and pairwise correlations, the maximum entropy distribution is
%
%
\begin{equation}
P(\{\sigma_{\rm i}\}) = \frac{1}{Z}e^{-\mathcal{H}}={1\over{Z}}\exp\left[
\sum_{{\rm i}=1}^N h_{\rm i}\sigma_{\rm i} + {1\over 2}\sum_{{\rm i}\neq {\rm j}}^N J_{\rm ij} \sigma_{\rm i}\sigma_{\rm j} \right] , \label{ham}
\end{equation}
where the ``magnetic fields'' $\{h_{\rm i}\}$ and the ``exchange couplings'' $\{J_{\rm ij}\}$ have to be set to reproduce the measured values of $\{\langle \sigma_{\rm i}\rangle\}$ and $\{C_{\rm ij}\}$.   This is {\em exactly} the Ising model with pairwise interactions among the spins.  Thus, the maximum entropy construction derives the Ising model from real data, rather than postulating it as an approximation to the underlying dynamics.  The construction is not an analogy or metaphor, but an exact mapping---Eq (\ref{ham}) should predict, given the measured correlations between pairs of neurons, the probability of all states for the whole network of $N$ neurons.  Further, this is a minimal model, in that the real network can have more structure than predicted by the maximum entropy model, but not less.

Conceptually, $J_{\rm ij}$  describes the direct mutual interaction between neurons $\rm i$ and $\rm j$ that remains after the contributions from other interactions in the network through more circuitous paths have been  disentangled from the corresponding correlation $C_{\rm ij}$, and $h_{\rm i}$ represents the neuron's intrinsic bias towards firing or silence \cite{schneidman+al_03,schneidman+al_06}.
This construction thus takes us from the experimentally accessible correlation functions $C_{\rm ij}$ and $\langle \sigma_{\rm i}\rangle$ to the underlying Hamiltonian  $\mathcal{H}$, which in turn determines the probability of every binary word $\{\sigma_{\rm i}\}$ in the neural codebook.  This path is the inverse of the usual problem in statistical physics, where we take the fields and couplings to be known (or chosen from a known distribution) and we must calculate the observable correlation functions.

The surprising result of Ref \cite{schneidman+al_06} is that the pairwise Ising model provides a very accurate description of the combinatorial patterns of spiking and silence, $\{\sigma_{\rm i}\}$, in ganglion cells of the salamander retina as they respond to natural and artificial movies, and in cortical cell cultures.
In other words, the frequency of appearance of all binary patterns across $N$ neurons can be explained by the interactions between pairs of neurons;  consequently, Eq (\ref{ham}) only requires  $O(N^2)$ parameters instead of the original $2^N$ to fully specify the distribution.  After the initial success in the salamander retina, similarly encouraging results were obtained in the primate retina, under very different stimulus conditions \cite{shlens+al_06,shlens+al_09},  in visual cortex \cite{cosyne,yu}, and in networks grown in vitro \cite{tangetal}.  Most of these detailed comparisons of theory and experiment were done for groups of $N\sim 10$ neurons, 
small enough that the full distribution $P_{\rm expt}(\{\sigma_{\rm i}\})$ could be sampled experimentally and used to assess the quality of the pairwise maximum entropy model of Eq (\ref{ham}).

We report here on our efforts to move towards larger networks of neurons and explore the kinds of collective effects that might be expected in networks of a few hundred elements in size. Since making the first version of our results available \cite{preprint}, this work has stimulated research into the tractability and approximate methods for solving the maximum entropy problem \cite{Broderick+al_08,monasson,nirenberg,roudi1,roudi2,roudi3} and generalizations of the pairwise model \cite{marre,bethge,thierry,cessac}. There is also growing interest in the use of maximum entropy methods to describe a wider range of biological systems, from protein structure to gene regulatory networks to ecosystems \cite{gtphd,ranganathan,seno,volkov,dhadialla,weigt}.
Here we present a more detailed exposition, arguing that our findings, in addition to providing a good fit to the data, can help us understand how networks of neurons function.

In detail, we  extend the maximum entropy results to $N=40$ neurons using Monte Carlo methods.  To motivate the discussion, we start by illustrating how strong collective effects can emerge in a toy mean-field Ising model used in the context of an inverse problem. We then extract realistic Ising models from the data; by studying subnetworks drawn from the full 40-neuron network, we first present an argument for why pairwise models can be successful in accounting for neural data, at least for small enough $N$. We  then argue that the observed subnetworks are typical of an \emph{ensemble}, for which we provide an explicit construction and out of which we can draw larger, synthetic networks. Remarkably, these larger networks seem to be poised very close to a critical point and exhibit a rich vocabulary of locally stable states, allowing us to hypothesize about possible combinatorial coding mechanisms and, as a result, predict two interesting collective behaviors of neural networks that should become visible in the next generation of experiments \cite{segev+al_04,gunning+al_05,dario}. 

\section{Distributions of output words}

As a concrete example, we consider  a population of retinal ganglion cells---the neurons which provide the brain with all of its input about the visual world---responding to naturalistic movies.  In such experiments, it is conventional to ask for a model which can predict the response of the neurons to arbitrary stimuli, $P(\{\sigma_{\rm i}\}|{\rm stimulus})$.  In the natural setting, stimuli are drawn from a space of very high dimensionality, and so constructing this map from stimuli to responses is very challenging \cite{fairhall_berry,keat,pillow,kolia}.
Alternatively, we can ask for a `dictionary' that describes the stimuli consistent with particular patterns of activity, $P({\rm stimulus}| \{\sigma_{\rm i}\})$ \cite{spikes}.   Here we take a very different approach, largely ignoring the visual stimulus and trying to understand the distribution of responses, $P(\{\sigma_{\rm i}\})$.  Before proceeding, we explain why this seemingly more limited problem is of interest.

Even when we measure the correlation between two neurons, $C_{\rm ij}$, the usual approach is to dissect the correlation into contributions which are intrinsic to the network and those which can be ascribed to common, stimulus driven inputs.  The idea of decomposing correlations  dates back to a time when it was hoped that correlations among spikes could be used to map the synaptic connections between neurons \cite{perkel+bullock_68}.  In fact, in a highly interconnected system, the dominant source of correlations between two neurons---even if they are entirely intrinsic to the network---will always be through the multitude of indirect paths involving other neurons \cite{ginzburg+sompolinsky_94}.
On the other hand, for neurons in the early stages of sensory processing, it is often suggested that the responses are ``conditionally independent'' given the stimulus, that is $P(\{\sigma_{\rm i}\}|{\rm stimulus}) = \prod_{\rm i} P_{\rm i}(\sigma_{\rm i}|{\rm stimulus})$ \cite{latham+nirenberg}.
In the particular case considered here, we know that, for populations of $N=10$ neurons, this model already fails dramatically \cite{schneidman+al_06}.

The question of whether correlations are driven by the stimulus or are intrinsic to the network is not a question that the brain can answer.  We, as external observers, can repeat the stimulus exactly, and search for correlations conditional on the stimulus, but this is not accessible to the organism.  The brain has access only to the output of the retina, patterns of activity which are drawn from the distribution $P(\{\sigma_{\rm i}\})$.   If the responses $\{\sigma_{\rm i}\}$ are codewords for the visual stimulus, then the entropy of this distribution sets the capacity of the code to carry information.  Word by word, $-\log P(\{\sigma_{\rm i}\})$ determines how surprised the brain should be by each particular pattern of response, including the possibility that the response was corrupted in transmission and thus should be corrected or ignored \cite{hopfield_08}.  In a very real sense, what the brain `sees' are sequences of states drawn from the distribution $P(\{\sigma_{\rm i}\})$.  In the same spirit that 
many groups have studied the statistical structures of natural scenes \cite{field, dong+atick, ruderman,simoncelli+schwartz,bethge,stephens}, we would like to understand the statistical structure of the codewords that represent these scenes.   We shall see that this structure has features which are suggestive of new ideas about the nature of the retinal code.

\section{A simple example} \label{sectoy}

We would like to illustrate these ideas with a simple example.  
Imagine that we record from $N$ neurons, and we find that all of them have the same mean rate of spiking, $\bar r$.  Further, if we look at any pair of neurons, the probability of both spiking in the same small window of time is a bit larger than expected if they were independent.  To be precise, we choose time windows of duration $\Delta\tau$,  so the probability of a spike is $q = \bar r \Delta \tau$ and the probability of coincident spikes from two particular cells 
 is $p_c = q^2 (1+\epsilon)$.
We want to describe this network as above, with Ising variables $\sigma_{\rm i} = +1$ for spiking and $\sigma_{\rm i} = -1$ for silence.  Then we have
\begin{eqnarray}
\langle \sigma_{\rm i}\rangle &=& -1 + 2q,\\
\langle \sigma_{\rm i}\sigma_{{\rm j}\neq {\rm i}}\rangle &=& \langle\sigma_{\rm i}\rangle^2 + 4\epsilon q^2 .
\end{eqnarray}
Since all neurons and pairs are equivalent, the maximum entropy model consistent with pairwise correlations has the simpler form,
\begin{equation}
P(\vec\sigma ) = {1\over {Z(h,J)}}\exp\left[ h \sum_{{\rm i}=1}^N \sigma_{\rm i} + {J\over 2}\sum_{{\rm i}\neq{\rm j}}^N  \sigma_{\rm i}\sigma_{\rm j}\right] ,
\end{equation}
which is just the mean field ferromagnet (assuming that $J$ is positive, and that the  temperature $k_B T$, which here does not have a direct physical interpretation, has been set to 1).   The solution of the mean field model is well known \cite{kadanoff}; here we recall a few details which will be especially important in the context of neurons.

As usual, everything we need to know is encoded in the partition function, which we evaluate by introducing an auxiliary field $\phi$:
\begin{eqnarray}
Z(h,J) &\equiv& \sum_{\{\sigma_{\rm i}\}} \exp\left[ h \sum_{{\rm i}=1}^N \sigma_{\rm i} + {J\over 2}\sum_{{\rm i}\neq{\rm j}}^N  \sigma_{\rm i}\sigma_{\rm j}\right]\\
&=& 2^N e^{-{NJ}/ 2}\int {{d\phi}\over{\sqrt{2\pi J}}} \exp\left[ - N F_N(\phi; h, J)\right] ,\nonumber\\
&&\label{mf2}
\end{eqnarray}
where the effective free energy is
\begin{equation}
F_N(\phi; h, J) = {{\phi^2}\over{2 NJ}} - \ln\cosh (h+\phi) .
\label{defF}
\end{equation}
If $N$ is large and we hold $NJ$ finite, then the integral in Eq (\ref{mf2}) is dominated by the mean field or classical value for $\phi$, defined by
the minimum of $F_N$,
\begin{equation}
\phi_c = NJ\tanh (h + \phi_c ) . \label{phic}
\end{equation}
Evaluating $Z$ to first order in the fluctuations around $\phi_c$, we have
\begin{widetext}
\begin{equation}
\ln Z(h,J) = N\left[ \ln 2 - {J\over 2} - {{\phi_c^2}\over{2 NJ}} + \ln\cosh (h+\phi_c ) \right] - {1\over 2}\ln \left[ 1 - NJ\left[(1 -\tanh^2(h+\phi_c))\right]\right] + \cdots ,
\end{equation}
where $\cdots$ refers to higher order terms in $1/N$, again assuming that $NJ$ stays finite as $N$ becomes large.

To fix the values of $h$ and $J$, we need to solve for the expectation values $\langle \sigma_{\rm i}\rangle$ and $\langle \sigma_{\rm i}\sigma_{\rm j}\rangle$:
\begin{eqnarray}
{1\over N} {{\partial \ln Z(h,J)}\over{\partial h}}
&=& 
\langle \sigma_{\rm i}\rangle = -1 + 2 q 
t- {1\over N} {{NJ t (1-t^2)}\over {\left[1- NJ(1-t^2)\right]^2}} ,\label{mfmean} \\
{{\partial\ln Z(h,J)}\over{\partial (NJ)}} &=& 
{1\over {2N}} {\Bigg\langle} \sum_{{\rm i}\neq{\rm j}}^N  \sigma_{\rm i}\sigma_{\rm j}{\Bigg\rangle} 
= {{(N-1)}\over 2} \left[ \langle\sigma_{\rm i}\rangle^2 + 4\epsilon q^2\right] \\
 &=& {N\over 2}t^2 - {1\over 2}
 +{1\over 2}{{1-t^2}\over{1- NJ(1-t^2)}} \left[ 1 - {{2NJt^2}
\over{1- NJ(1-t^2)}}\right], \label{mvcov}
\end{eqnarray}
\end{widetext}
where $t\equiv \tanh(h+\phi_c)$.
Rearranging and retaining terms to order $1/N$ or larger, we have
\begin{eqnarray}
4\epsilon q^2=\frac{1}{N}\frac{NJ (1-t^2)^2}{1-NJ(1-t^2)}. \label{corscale}
\end{eqnarray}
We recall that we usually require $NJ$ to scale as $NJ \sim {\cal O}(1)$ at large $N$, because we want to have a well defined thermodynamic limit, in which energy and entropy are extensive quantities.  In particular, with $NJ = J_0$, we can write the effective free energy in Eq (\ref{defF}) as  $F_N(\phi; h, J) = F(\phi; h,J_0)$, where $F$ has no explicit $N$ dependence.  
Another key (and familiar) point is that, with the usual scaling of $NJ$ constant as $N$ becomes large,  the correlations behave as $\epsilon \propto 1/N$ [Eq (\ref{corscale})], and in the $N\rightarrow \infty$ limit, they must decay to zero.

To connect the Ising models to the data, we face a new problem: we do not have direct access to the parameters $J$ and $h$ but rather have to infer them from measured $q$ and $\epsilon$. As we examine recordings of increasingly large subsets of neurons taken from a dense patch of the retina, we could find that instead of decaying to 0, the average pairwise correlation $\epsilon$ stays constant. This would mean that $NJ$ cannot be constant as we consider larger groups of neurons, and that hence increasingly large subsets of neurons do not comprise an ensemble with a conventional thermodynamic limit. 

It is easy to find an example of such a behavior and gain intuition if we limit ourselves to a (somewhat unrealistic) case when $h=0$. Then, Eq (\ref{phic}) for the mean field value of $\phi_c$ has either 1 or 3 solutions, depending on the value of $J$ -- the solution $\phi_c=0$ corresponds to the paramagnetic phase, and the non-zero solutions correspond to spontaneous magnetization. The transition from 1 to 3 solutions is the phase transition in the Ising ferromagnet, and it occurs at the critical value $J_0^*=1$ in the thermodynamic limit.

Suppose that we are in the paramagnetic phase, with the solution $\phi_c=0$, $t=0$ and $J_0 < 1$. Equation (\ref{mfmean}) then tells us that $\langle \sigma_{\rm i}\rangle = 0$, or $q=0.5$. If the average correlation is found experimentally to be $\epsilon$, then we can compute how $J_0=NJ$ must scale from Eq (\ref{corscale}):
\begin{equation}
J_0(\langle \sigma_i\rangle = 0, \epsilon)=\frac{N\epsilon}{1+N\epsilon}.
\end{equation}

As we assumed, for every finite $N$, $J_0$ will stay below the critical value of 1 and thus choosing the solution with $t=0$ is self-consistent. We also note, however, that as $N$ increases, the system approaches the critical value $J_0=1$, regardless of the value of $\epsilon$, as long as that does not decrease faster than $1/N$. To detect such an onset of criticality, one could perturb the coupling constans $h$ and $J$ by introducing a fictitious temperature $T$, perturbing it around the value 1, and observing the emergence of the peak in heat capacity $C(T)=T\partial S/\partial T$ at $T=1$. 

In the real data the mean firing rates and the correlations are not homogenous across the population, and the $\langle \sigma_{\rm i}\rangle=0$ assumption clearly is not valid, because the neurons spike infrequently, making $q\ll 1$, so that our toy model cannot be taken too literarily.
Nonetheless, this exercise teaches us that one must be careful in transferring our intuition about thermodynamic limits to the case of maximum entropy models for real networks.  It is not just that real networks have finite $N$; indeed, many networks probably have $N$ sufficiently large that some sort of large $N$ approximation is valid.  The problem is rather that, since the correlations among pairs of neurons are experimentally measurable, we are not free to let these vary as we imagine recording from more and more neurons.
More precisely, having a non--zero correlation between all pairs in a large, homogeneous network is inconsistent with the large $N$ limit except at the critical point.  This prepares us for the possibility that, even in a more realistic, inhomogeneous system, the common observation of nontrivial correlations among most pairs of neurons points toward an interesting regime of operation.


\section{Learning maximum entropy models from real data}

We recall that maximum entropy models are the least structured, or most random, models consistent with known expectation values \cite{jaynes_57,schneidman+al_03}. The Ising model of Eq (\ref{ham})
thus  is  the minimal model forced upon us by measurements of mean spike probabilities and pairwise correlations. In maximum entropy models that match only  the measured mean spike probability but not pairwise correlations (the so-called \emph{independent models}), the couplings $J_{\rm ij}$ are zero, and the probability distribution factors into independent contributions from each neuron. As has been shown elsewhere, and as we also reiterate later in this paper, independent models completely fail to account for the data, and the pairwise extension is therefore the next minimal step that we are required to take. In addition, the pairwise model also is a minimal model in which we can expect to observe any interesting collective behaviors.

\begin{figure} 
   \centering
   \includegraphics[width=3.5in]{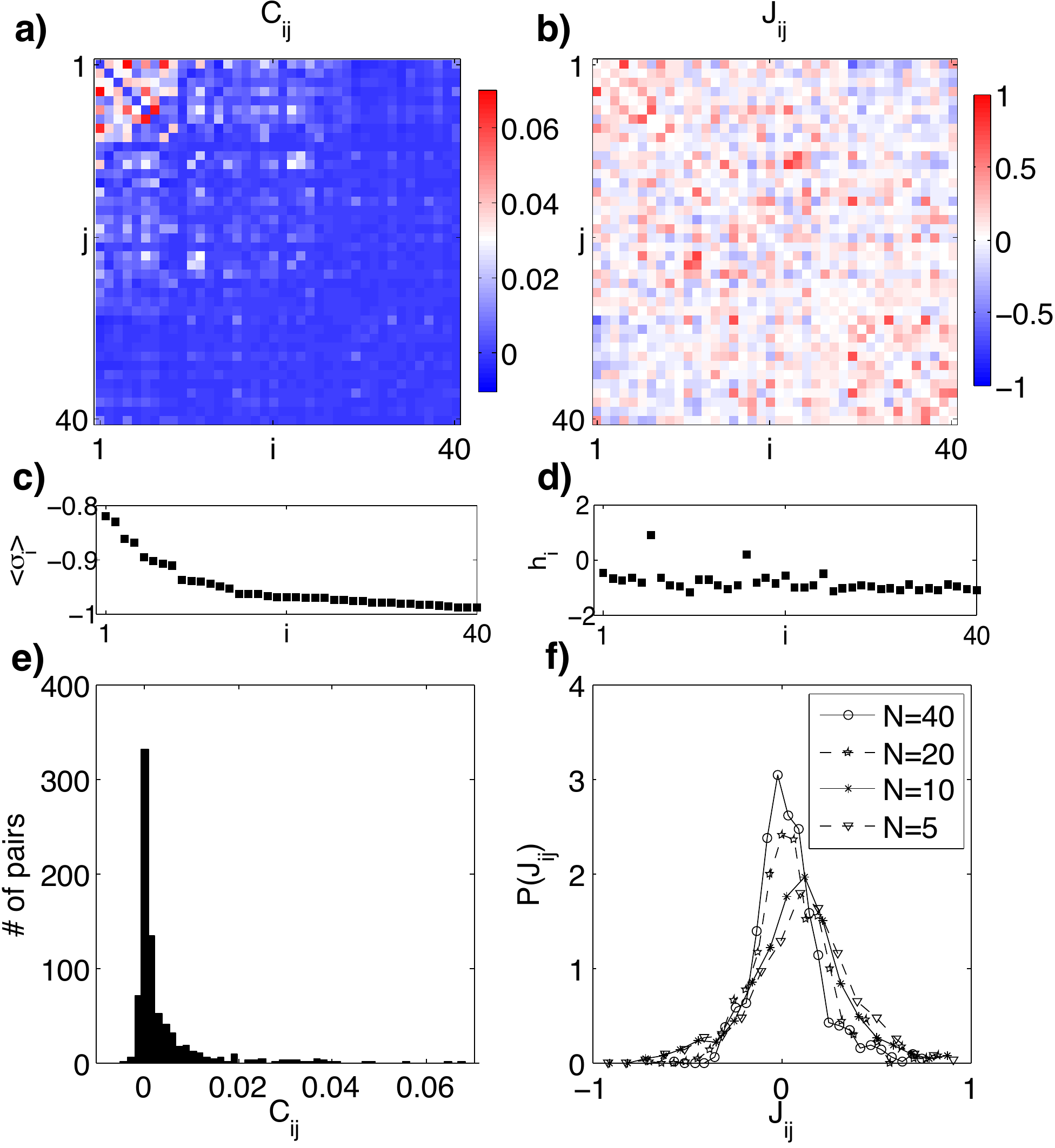}
   \caption{Expectation values and interactions.  At left, the expectation values computed from the experimental data; at right, the corresponding terms in the effective Hamiltonian of the maximum entropy model.  Neurons are numbered in order of decreasing mean spike rate.  (a) The pairwise connected  correlations, $C_{\rm ij} = \langle \sigma_{\rm i}\sigma_{\rm j} \rangle - \langle \sigma_{\rm i}\rangle\langle\sigma_{\rm j} \rangle$; note that we see both positive and negative correlations, and that these correlations are small.  
   (b) The interactions $J_{\rm ij}$ in the Ising model; note that interactions spread more uniformly through the network than the correlations.
   (c) The mean ``magnetization'' of the individual cells. 
   (d) The ``magnetic fields'' that express the intrinsic tendency of the neurons toward spiking or silence. (e) The histogram of correlations. (f) The distribution of interactions inferred for sub--networks of different size.
   }
   \label{fx}
   \end{figure}

To be concrete, we consider the salamander retina responding to naturalistic movie clips, as in the experiments of Refs \cite{schneidman+al_06,puchalla+al_05}.  The visual stimulus consists of a $26.2\e{s}$ movie that was projected onto the retina 145 times in succession, for a total of roughly one hour of stable recordings.
Using bins of $\Delta \tau=20\e{ms}$ along the time axis yields 1310 samples per movie repeat, for a total of 189950 binary word samples, where in each time bin each neuron can either fire or be silent. The effective number of independent samples is smaller because of correlations across time; using bootstrap error analysis we estimate $N_{\rm samp} \sim 7\E{4}$ \cite{gtphd}. Under these conditions, pairs of cells within  $\sim 200\,\mu{\rm m}$ of each other have correlations drawn from a homogeneous distribution; the correlations decline at larger distance \cite{segev_physiol}.   This correlated patch contains $N\sim 200$ neurons \cite{segev+al_04}, of which we record from $N=40$  \cite{schneidman+al_06}.
The values of $\langle \sigma_{\rm i}\rangle$ and $C_{\rm ij}$ for this population of cells are shown in Fig \ref{fx}a and \ref{fx}c, respectively.

The central problem is to find the magnetic fields and exchange interactions that reproduce the observed pairwise correlations.  It is convenient to think of this problem more generally:  We have a set of operators $\hat O_\mu (\{\sigma_{\rm i}\})$ ($=\left\{\sigma_{\rm i},\sigma_{\rm i}\sigma_{\rm j}\right\}$ for the pairwise model) on the state of the system, and we consider a class of models 
\begin{eqnarray}
P(\{\sigma_{\rm i}\}| {\vek g}) &=& {1\over{Z({\vek g})}}
\exp\left[ \sum_{\mu =1}^K g_\mu \hat  O_\mu (\{\sigma_{\rm i}\}) \right],\\
Z({\vek g}) &=& \sum_{\{\sigma_{\rm i}\}} \exp\left[ \sum_{\mu =1}^K g_\mu \hat  O_\mu (\{\sigma_{\rm i}\}) \right];
\end{eqnarray}
our problem is to find the couplings $\vek g$ ($\equiv\left\{h_{\rm i},J_{\rm ij}\right\}$ for the pairwise model) that generate the correct expectation values, which is equivalent to solving the equations
\begin{equation}
{{{\partial \ln Z({\vek g})} \over {\partial g_\mu}}} = \langle \hat O_\mu (\{\sigma_{\rm i}\}) \rangle_{\rm expt} .
\end{equation}
Up to $N\sim20$ cells we can solve  exactly, but
this approach does not scale to $N=40$ and beyond because the partition sum $Z$ contains a number of terms that is exponential in $N$.  For larger systems, this ``inverse Ising problem'' or Boltzmann machine learning, as it is known in computer science \cite{Hinton}, is a hard computational problem that has to be solved by approximate schemes \cite{Broderick+al_08,monasson,thierry}.

Given a set of coupling constants $\vek g$, we can estimate the expectation values 
$\langle\hat O_\mu \rangle_{\vek g}$ by Monte Carlo simulation.  Increasing the  coupling $g_\mu$ will increase the expectation value $\langle\hat O_\mu \rangle$, so a plausible algorithm for learning   $\vek g$ is to increase each $g_\mu$ in proportion to the deviation of $\langle\hat O_\mu \rangle$ (as estimated by Monte Carlo) from its target value (as estimated from experiment).  This is not a true gradient ascent, since changing $g_\mu$ has an impact on operators $\langle\hat O_{\nu\neq\mu} \rangle$, but such an iteration scheme does have the correct fixed points; heuristic improvements 
include a slowing of the learning rate with time and the addition of some `inertia', so that we update $g_\mu$ according to 
\begin{equation}
\Delta g_{\mu}(t+1) = - \eta(t) \left[
\langle\hat{O}_\mu\rangle_{{\vek g}(t)} 
-
\langle\hat{O}_\mu  \rangle_{\rm expt} 
\right]
+\alpha \Delta g_{\mu}(t),
\label{eq_ising_learn}
\end{equation}
where $\eta(t)$ is the time--dependent learning rate and $\alpha$ measures the strength of the inertial term.

\begin{figure}[b] 
   \centering
   \includegraphics[width=3.5in]{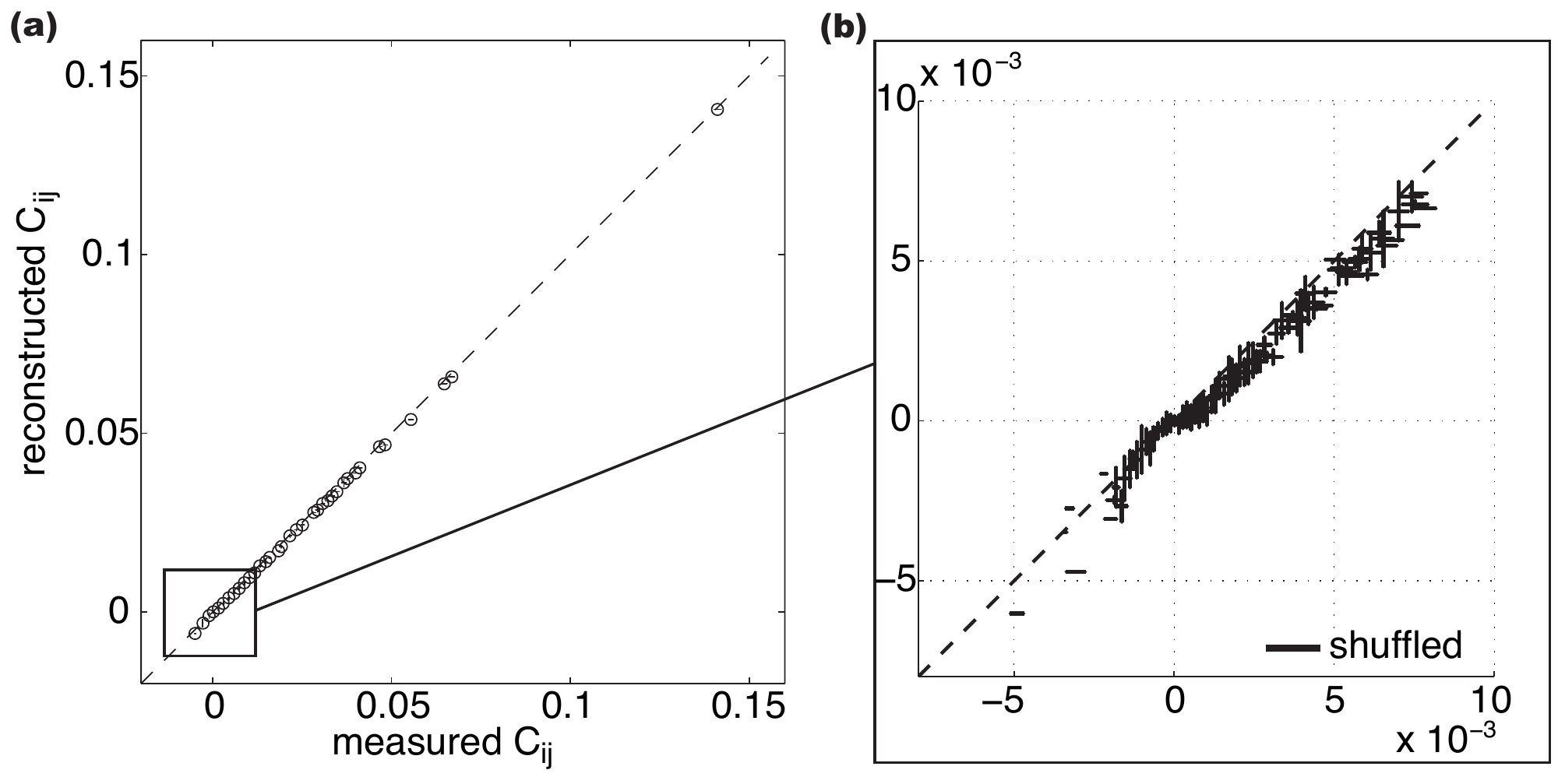}
   \caption{{\bf a)} Precision of the Ising model for $N=40$ neurons learned via  Eq (\ref{eq_ising_learn}): measured covariances are binned  on the x-axis and plotted against the corresponding reconstructed covariances on the y-axis; vertical error bars are the standard deviation within the bin, horizontal error bars are bootstrap errors on covariance estimates.  {\bf b)} Zoom--in for small $C_{\rm ij}$.  The scale bar represents the standard deviation  of $C_{\rm ij}$ in shuffled data, where all observed correlations arise from insufficient sampling. }
   \label{f1}
   \end{figure}

We used different variations on this basic strategy for networks of different size.  As noted above, for $N\leq 20$ we solve exactly using a custom implementation of algorithm described in Ref \cite{dudik}; this involves fully evaluating the partition sum.
For $N=40$, we first obtained a good initial approximation for $\vek{g}$ by contrastive divergence Monte Carlo \cite{cd}.  Then we followed up until convergence by using Eq (\ref{eq_ising_learn}) directly and decreased the learning rate $\eta(t)$ as $O(1/t)$ or slower according to a custom schedule, and kept the inertia $\alpha$  at zero. 
Later in the paper we consider synthetic networks with $N=120$, and for these we use $\alpha = 0.9$. 
 Learning the couplings $\vek g$ was slow for $N=120$ synthetic networks, but eventually converged; even in this case $C_{\rm ij}$ converged to within $10\%$ for the largest quartile of elements by absolute value,
and within $15\%$ for the largest half,
without obvious systematic biases.  In addition to these algorithmic issues, the Hamiltonian was rewritten such that $J_{\rm ij}$ was constraining $(\sigma_{\rm i}-\langle\sigma_{\rm i}\rangle_{\rm expt})(\sigma_{\rm j}-\langle\sigma_{\rm j}\rangle_{\rm expt})$, and we found that this removed   biases in the reconstructed covariances.

Figure \ref{f1} shows the success of the learning algorithm by comparing the measured pairwise correlations to those computed from the inferred Ising model for 40 neurons. The reconstructions of the means, $\langle\sigma_{\rm i}\rangle$, are not shown, but are accurate to better than $1\%$.

The pairwise maximum entropy model has many fewer parameters ($820$) than there are possible states of the network ($\sim 10^{12}$); more importantly, the effective number of independent sample in our data set is $\sim 7\times 10^4$, and this also is much larger than the number of parameters.  Further, the parameters of the model are functions of the mean firing rates and pairwise correlations.  Thus, to the extent that we can measure these quantities reliably, there is no issue of having a model which is too complex to be supported by the data.  Nonetheless, one would like an independent test to show that we are not `over--fitting' to a limited data set.  To do this, we divide the available data in half to create separate training and testing data.  We then use the model learned from the training data to compute the average log--likelihood  of the data in each half, and measure the difference  
\begin{eqnarray}
\Delta (k) &=&  - {\bigg\langle}  \ln P(\{\sigma_{\rm i}\} | \vek{g}_k) {\bigg\rangle}_{\{\sigma_{\rm i}\}\in \rm test\,\, data} \nonumber\\
&&\,\,\,\,\,\,\,\,\,\, +  {\bigg\langle}  \ln P(\{\sigma_{\rm i}\} | \vek{g}_k) {\bigg\rangle}_{\{\sigma_{\rm i}\}\in \rm training\,\, data} ,
\end{eqnarray}
where $k$ indexes one random split of the data into training and test halves, and $\vek{g}_k$ is the set of parameters that we learn from the training data on that split.    It should be noted that in the computation of $\Delta$, the partition function $Z(\vek{g}_k )$ cancels, so really we are computing the difference in the mean ``energy'' between the training and test data.    With twenty random choices for the training/test split, we find that $\Delta = 0.0078 \pm 0.0086$.  
Thus $\Delta$ has enough variance to consistently include zero with high probability, implying that over--fitting is not a problem. 


\section{Does the model `work'?}

\begin{figure} 
   \centering
   \includegraphics[width=3.4in]{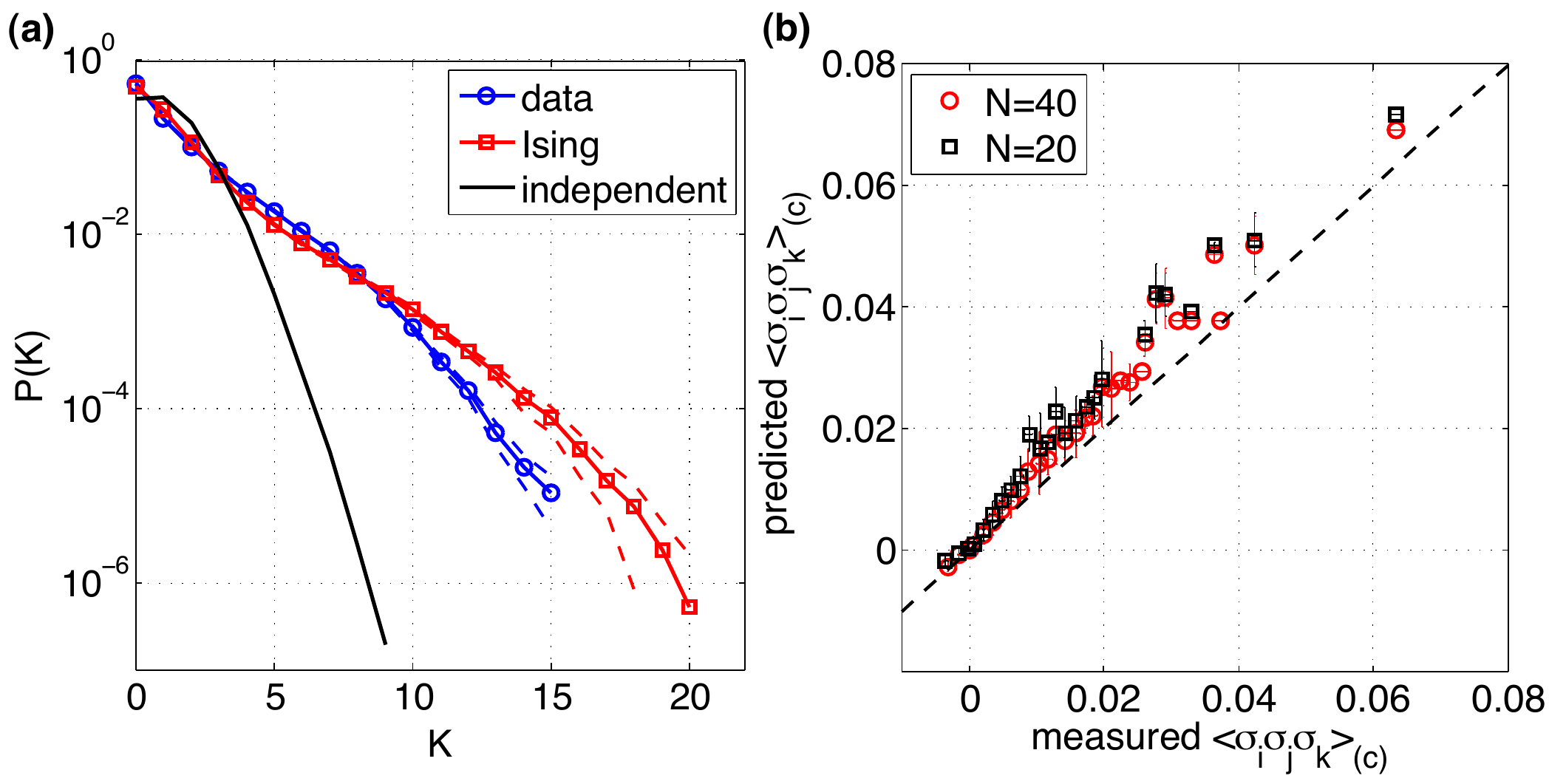}
   \caption{{\bf a)} Probability of observing $K$ simultaneous spikes (blue -- data, red -- Ising model), compared to the failure of the independent model, where $J_{\rm ij}=0$ (black). Dashed lines show error estimates. {\bf b)}  Measured vs predicted connected three--point correlations for 40 neurons (red) and for an exact Ising model for a 20 neuron subset (black). Horizontal error bars are bootstrap estimates from data, vertical error bars are computed across 20 Monte Carlo runs.} 
 \label{lf2}
\end{figure}

For $N=10$ or even $N=20$, we can compare the predictions of the maximum entropy model with the observed frequencies of the network states $\{\sigma_{\rm i}\}$.  For $N=40$, this is no longer possible, since the number of states is much larger than the number of samples available in the experiment.  What we can do is to compute, in the maximum entropy model, various statistics beyond the pairwise correlations, and check these quantities against estimates from the experimental data.  In Fig \ref{lf2} we show two examples of this approach. 

First we calculate the probability that $K$ of the $N$ neurons spike in the same time bin.  The data show that $P(K)$ is an approximately exponential distribution, enormously far from the roughly binomial distribution expected if the neurons were spiking independently.  The exponential structure of $P(K)$ is well reproduced by the predictions from the maximum entropy model; the maximum entropy model has a tail which is slightly too heavy.  In more detail, the Ising model underestimates the probability of the no--spike pattern by a few percent, $P_{\rm expt}(K=0) =0.550$ vs. $P_{\rm Ising} (K=0) =0.502$. The same deviation is already observed at $N=20$, where $P_{\rm expt}(0) =0.621$ vs. $P_{\rm Ising} (0) =0.599$, and since at this network size we reconstruct the Ising model using an exact algorithm, the deviations at $N=40$ are not due to limitations of our Monte Carlo algorithm. Because the no--spike pattern is sampled well in our dataset, this deviation is significant and indicates that higher order interactions are starting to have an effect, albeit a small one.

As a second test, we consider the correlations among triplets of neurons.  
We look for the correlated deviations of triplets of neurons from their mean firing rates and define the connected third-order correlation coefficients, $\langle\sigma_{\rm i}\sigma_{\rm j}\sigma_{\rm k}\rangle_{(c)}=\langle(\sigma_{\rm i}-\langle\sigma_{\rm i}\rangle)(\sigma_{\rm j}-\langle\sigma_{\rm j}\rangle)(\sigma_{\rm k}-\langle\sigma_{\rm k}\rangle)\rangle$. 
Of course the independent model predicts that these will be zero, whereas about a third of the $(40)(39)(38)/(3!) \sim 10^4$ distinct triplets of neurons have correlations which are significantly different from zero.  The maximum entropy model certainly captures the trend of these correlations, although it does tend to overestimate the large ones slightly, by $\sim 7\%$ on average.  It is worth recalling that these $\sim 10^4$ triplet correlations are being predicted from a model that is determined only by the $820$ measured means and pairwise correlations, with no additional fitting allowed.  In this context, $7\%$ errors are small, and are detectable only because we have tens of thousands of samples.

The maximum entropy formalism is a series of ever better approximations to the true distribution $P_{\rm expt}(\left\{\sigma_i\right\})$, in which the approximations match the data to increasing correlation orders. There is no a priori reason why pairwise order alone should suffice, but our results show that even at $N=40$ the Ising model performs exceedingly well, closing most of the gap between  the independent model and the real data.   The correct view of this result is not that pairwise model is exactly correct, since clearly it isn't, but rather that this model is capturing a large fraction of the correlation structure in the data, even the higher order structure.

Given that the pairwise model does work very well, there are at least three questions we could ask.  One is to address the small departures from the model, either by adding explicit higher order interactions, or by trying to infer the existence of `hidden' elements that generate effective multi--neuron interactions even if the dominant elementary interactions are among pairs.  This is a question that we leave for future work.  The next question is to try and understand why the model works as well as it does.  Finally, given that the model works, we can ask what it teaches us about the network as a whole, and about the neural code in the retina.  We take up these two questions in the next sections.


%

\section{Exploring  subnetworks}

In the current experimental setup, simultaneous recordings can only be made on a subset of neurons from a patch of the retina with highly correlated firing activity. It is thus relevant to ask about the effects of unobserved or `hidden' nodes on the validity of the pairwise model. 
To explore this issue with the data we do have, we can `hide' some of the cells in the $N=40$ network to which we have access, and ask how this changes the quality of the maximum entropy description. Intuitively,  it is surprising that pairwise models work well both on $N=40$ neurons and networks consisting of subsets of these neurons, as in the original Ref \cite{schneidman+al_06}. We would expect that not observing $\sigma_\chi$ will induce a triplet interaction $J_{\alpha\beta\gamma}$ among neurons $\left\{\sigma_\alpha,\sigma_\beta,\sigma_\gamma\right\}$ for any triplet in which there were pairwise couplings between $\sigma_\chi$ and all triplet members. 
Additionally, by comparing the parameters in the full network, $\vek{g}^{(40)}$, with their corresponding averages from different subnets of size $N=20$, $\vek{g}^{(20)}$, we find that  the couplings $J_{\rm ij}$ are left almost unchanged, while magnetic fields $h_{\rm i}$ change substantially.  

To explain both phenomena, we 
examine the flow of the couplings under decimation.  Let us start by including a three--body interaction term $J_{\rm ijk}\sigma_{\rm i}\sigma_{\rm j}\sigma_{\rm k}$ into the $N$-neuron Hamiltonian $\mathcal{H}^{(N)}$ of Eq (\ref{ham}) for $P^{(N)}(\left\{\sigma_{\rm i}\right\})$. Then, we isolate terms related to spin $\sigma_{\rm N}$, and marginalize over $\sigma_{\rm N}$ to obtain $P^{(N-1)}(\left\{\sigma_{\rm i}\right\})$. This summation results in
\begin{eqnarray}
P^{(N-1)}(\left\{\sigma_{\rm i}\right\})&\propto& e^{-\mathcal{H}^{(N-1)}} \times \label{expansion}\\
&&e^{\log\left(2\cosh\left[h_{\rm N}+\sum_{\rm i}J_{\rm iN}\sigma_{\rm i}+\frac{1}{2}\sum_{\rm ij}J_{\rm ijN}\sigma_{\rm i}\sigma_{\rm j}\right]\right)}. \nonumber
\end{eqnarray}
Because of the last term, the resulting expression for $P^{(N-1)}$ clearly does not have the same form as that for $P^{(N)}$. We can, however, expand around $\sigma_{\rm N}=-1$ up to terms of order $O((1+\sigma)^4)$ as long as $\sum_{\rm i} J_{\rm iN}(\sigma_{\rm i}+1)$, and the similar term for triplets, are small; we then look for a decimated Hamiltonian, $\mathcal{H}^{(N-1)}$, with renormalized couplings and a matching power series.  This results in the following flow of coupling when a single neuron $\sigma_{\rm N}$ has been marginalized over:
\begin{eqnarray}
h_{\rm i}&\rightarrow & h_{\rm i} + \omega \tilde{J}_{\rm iN} +\textstyle \sum_{\rm j} \beta_{\rm ij}+{\cal O}(\gamma,\delta),	\label{eq_dec1}\\
J_{\rm ij}&\rightarrow & J_{\rm ij}+\beta_{\rm ij}+{\cal O}(\gamma,\delta),\label{eq_dec2} \\
J_{\rm ijk} &\rightarrow& J_{\rm ijk} + {\cal O}(\gamma,\delta) \label{eq_dec3}
\end{eqnarray}
where $\tilde{J}_{\rm iN}=J_{\rm iN}-\sum_{\rm j} J_{\rm ijN}$, $\beta_{\rm ij} = \tilde{J}_{\rm iN}\tilde{J}_{\rm jN}(1-\omega^2)+\omega J_{\rm ijN}$ and  $\omega=\tanh(h_{\rm N}-\sum_{\rm i} J_{\rm iN}+\frac{1}{2}\sum_{\rm ij} J_{\rm ijN})$.  The terms  $\gamma,\delta \propto (1-\omega^2)$ originate from terms with 3 and 4 factors of $\sigma$, respectively.
The key point is that neurons spike very infrequently (on average in $\sim 2.4\%$ of the bins) and so $\langle \sigma_{\rm i}\rangle\approx-1$, in which case $\omega$ is approximately the hyperbolic tangent of the mean field at site $\rm N$ and is close to $-1$. If pairwise Ising  is a good model at size $N$, and couplings are small enough to permit expansion, then the corrections to pairwise terms after decimation, as well as $J_{\rm ijk}$, are multiplied by $1-\omega^2\ll 1$.  Sparsity of spikes therefore keeps the complexity in check: when a neuron cannot be observed, the correction to the pairwise couplings among the  `visible' neurons, as well as the appearance of higher order interactions between them, are suppressed.

We test these ideas by selecting 100 random subgroups of 10 neurons out of  20; for each, we compute the exact pairwise Ising model from the data, as well as 
applying 
Eqs (\ref{eq_dec1}--\ref{eq_dec3}), with full expressions for $\gamma$ and $\delta$,
10 times in succession to decimate the network from 20 cells down to the chosen 10. The resulting three--body interactions $J_{\rm ijk}$ have a mean and standard deviation ten times smaller than the pairwise $J_{\rm ij}$.  If we ignore these terms,  the average Jensen--Shannon divergence \cite{lin_91} between this probability distribution and the best pairwise model for the $N=10$  subgroups is $\overline{D}_{JS}= 9.3\pm 5.4 \times 10^{-4}\,{\rm bits}$,  smaller than the average divergence between either model and the experimental data, and means that $\gg10^3$  samples would be required to distinguish reliably between the two models.  At least for the range of $N$ examined here, the decimation calculation provides a useful approximation.

\begin{figure}[b]
   \centering
   \includegraphics[width=3.4in]{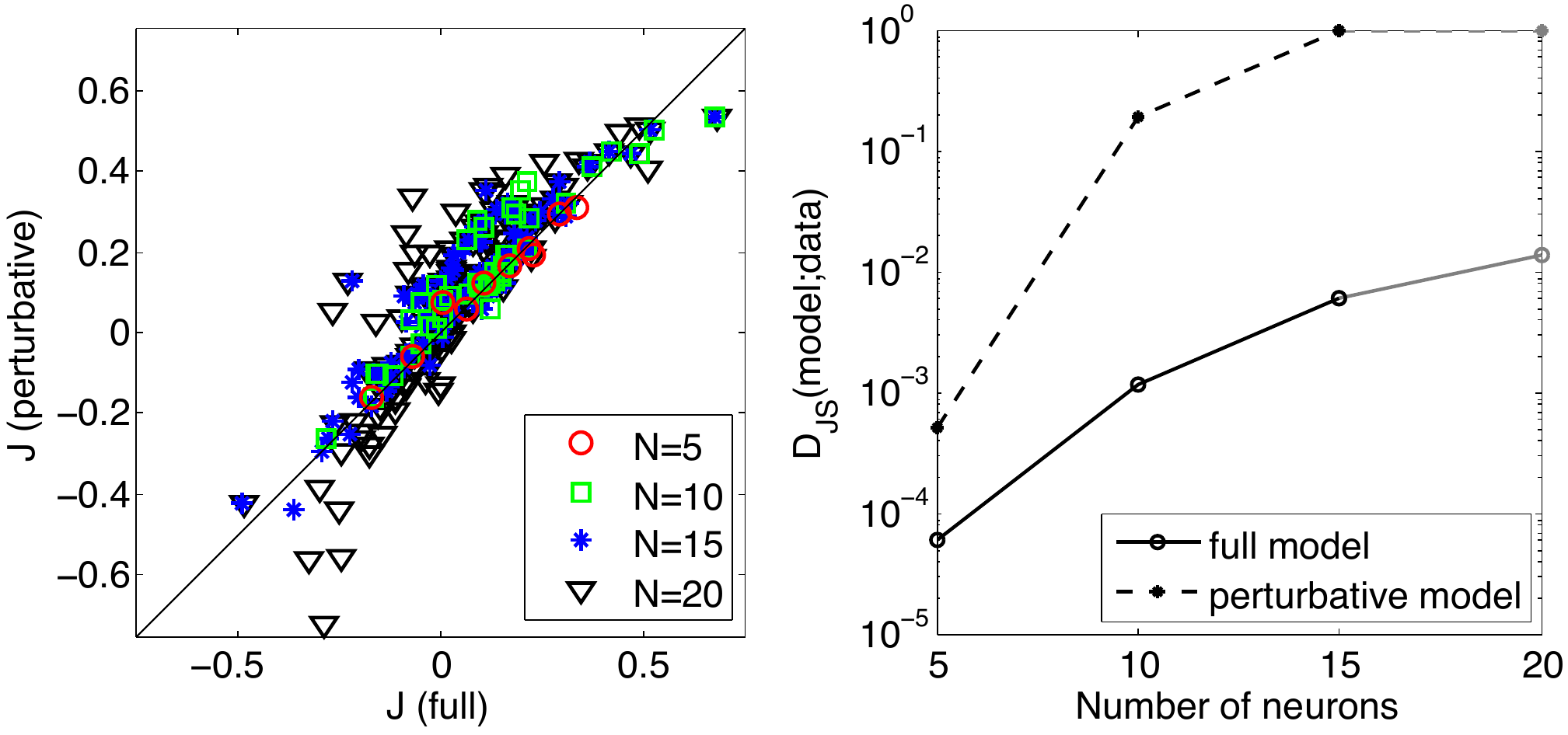}
   \caption{{\bf a)} The comparison of couplings $J_{\rm ij}$ for a group of $N=5,10,15,20$ neurons,  computed using the exact maximum entropy reconstruction algorithm, with the lowest order perturbation theory result, $J_{\rm ij}=\frac{1}{4}\log c_{\rm ij}$, where $c_{\rm ij}=\langle\tilde{\sigma}_{\rm i}\tilde{\sigma}_{\rm j}\rangle/(\langle\tilde{\sigma}_{\rm i}\rangle\langle\tilde{\sigma}_{\rm j}\rangle)$ and $\tilde{\sigma}_{\rm i}=0.5(1+\sigma_{\rm i})$ \cite{roudi2,sessak}. In the case of larger networks, the perturbative $J_{\rm ij}$ deviate more and more from equality (black line). {\bf b)} The exact Ising approximation, $P(\left\{\sigma_{\rm i}\right\})$, can be compared to the  distribution $P_{\rm expt}(\left\{\sigma_{\rm i}\right\})$, sampled from data; the solid line shows the Jensen-Shannon divergence between the two distributions, for four example networks of size $N=5,10,15,20$. The dashed line shows the same comparison in which the Ising model parameters, $\vek{g}=\left\{h_{\rm i},J_{\rm ij}\right\}$, were calculated perturbatively. Already at $N=10$ neurons, the perturbative model has an error two orders of magnitude larger than the exact Ising model. The point at $N=20$ shown in gray because $P(\left\{\sigma_{\rm i}\right\})$ cannot be reliably sampled from data.}
   \label{f5}
   \end{figure}

The sparsity of spikes  can explain the dominance of pairwise interactions: the higher order terms are not intrinsically small, but the fact that spiking is rare means that they do not have much chance to contribute.  Thus, the nature of the pairwise model is more like a Mayer cluster or virial expansion than like simple perturbation theory.    Of course, with finite $N$, all quantities must be analytic functions of the coupling constants, and so we expect that, if carried to sufficiently high order, any perturbative scheme will converge---although this convergence may become much slower at larger $N$, signaling genuinely collective behavior in large networks.  There are a number of reasons to think that (contrary to the suggestion in Ref \cite{roudi2}, but consistent with the conclusions of Ref \cite{monasson}) the real system we are studying is already outside the regime in which low order perturbative approaches are sufficient.  First, the simple relation between $J_{\rm ij}$ and $C_{\rm ij}$ predicted by the lowest order of perturbation theory \cite{roudi2, sessak} is violated, as shown in Fig \ref{f5}a, and the resulting errors in the predicted $P(\{\sigma_{\rm i}\})$ are significant, as shown in Fig \ref{f5}b.  Second, if we make a perturbative expansion of the entropy itself in powers of the measured correlations, even carrying this out to fourth order fails to work for $N>10$ neurons \cite{azhar_thesis}.
Finally, the standard deviation of `effective field' $\phi_{\rm i} = \sum_{\rm j} J_{\rm ij}\sigma_j$ that represents the influence on neuron $\rm i$ by its neighbors
is comparable to the intrinsic bias $h_{\rm i}$  for $N=40$. 

The question of whether pairwise models remain good for $N$ beyond 40, even in the regime when the above perturbation arguments break down, cannot be answered without further experimental data. We can conclude, however, that pairwise models are valid beyond the point where pairwise interactions $J_{\rm ij}$ merely represent trivial perturbative corrections to the dominant intrinsic biases $h_{\rm i}$, which happens already at $N\sim 10$.

\section{Network ensembles}  \label{secens}

In trying to develop a theoretical approach to biological systems, there is a tension between the search for universals and the need to engage with the details of specific systems.  In the present context, even if the maximum entropy models provide a perfect description of the probability distribution of network activity, one may worry that what we have learned is relevant only to the particular 40 neurons we happened to record from in this one experiment.  How can we generalize from these results?    One idea is that what we observe in this one experiment is typical of what we would find by drawing networks at random out of some ensemble of networks.  Our goal in this section is to identify this ensemble.


We start our search for meaningful ensembles of networks by characterizing the ``thermodynamics'' of the networks that we have observed.  Having constructed a maximum entropy model with some parameters $\vek{g}$, we can take this model seriously as a statistical mechanics problem and ask what happens as we change the ``temperature,'' which is equivalent to scaling all of the coupling 
${\vek g} \rightarrow {\vek g}/T$.  Notice that this is just one slice through the large space of parameters, but it is one which has a physical interpretation even though the temperature is not itself physical.  In particular we can define the entropy, heat capacity, and other thermodynamic variables.  


 \begin{figure} 
   \centering
   \includegraphics[width=3.3in]{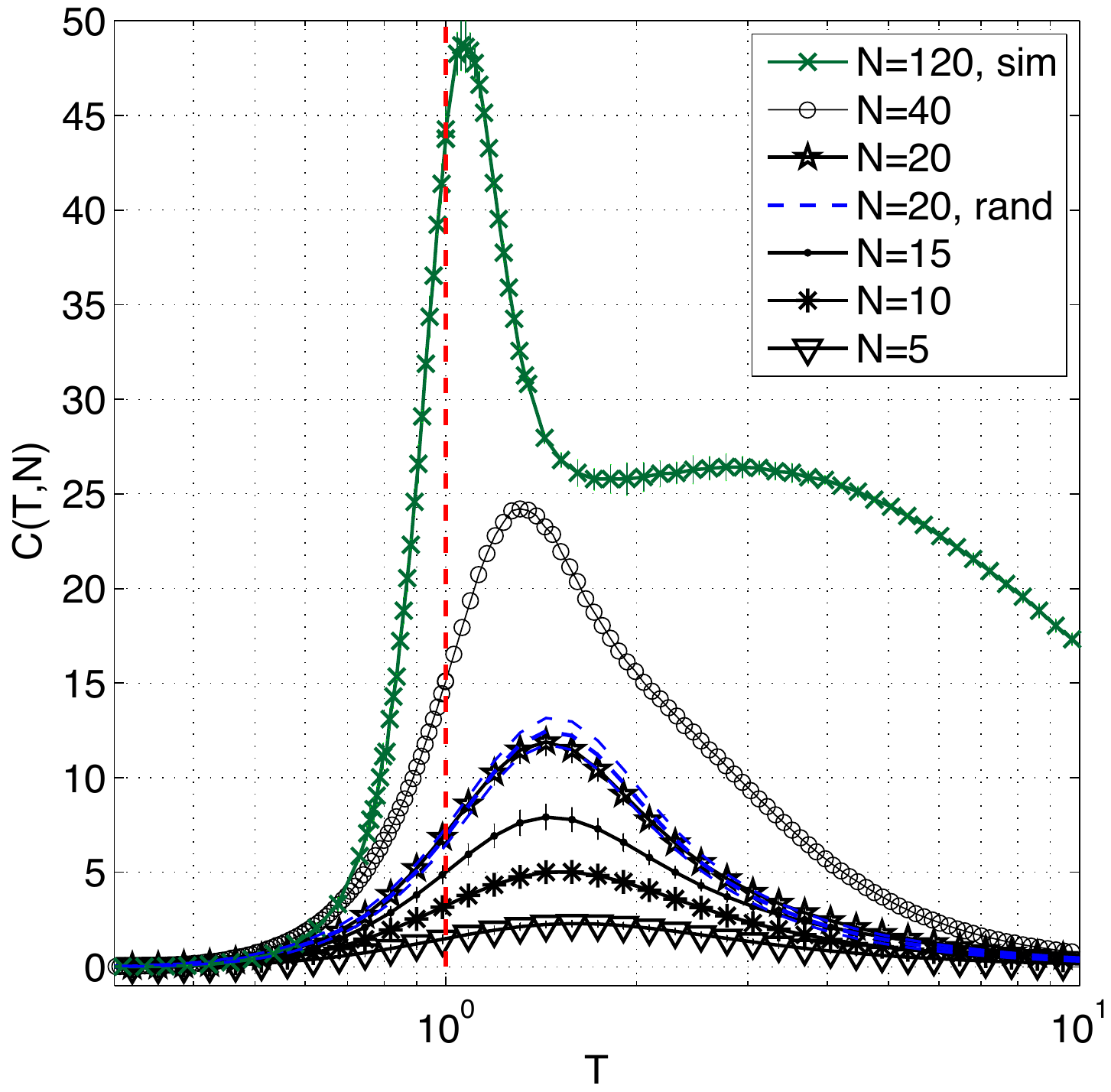}
   \caption{The heat capacity, $C(T,N)$, for systems of different sizes $N$. Ising models were constructed for 400 subnetworks of size 5, 180 of size 10, 90 of size 15 and 20, 1 full network of size 40 (all from data, black), and 3 synthetic networks of size 120 (green); vertical error bars are standard deviations across these examples. For networks of $N\leq 20$, the partition function (and therefore heat capacity) can be computed exactly; for $N>20$, we estimate heat capacity from the variance in energy during Monte Carlo sampling, i.e. $C(T,N)=\langle (\delta E)^2 \rangle/T^2$. The mean  and 1-sigma envelope for $C(T, N)$ of Ising models of randomized networks at $N=20$ are shown in blue dashed lines. We note that the peak of the heat capacity moves towards the operating point at $T=1$ with increasing size.
   } 
   \label{f2}
\end{figure}

To proceed, we first plot the dependence of heat capacity on temperature at various system sizes $N$ in   Fig \ref{f2}.  
The behavior of the heat capacity as a function of temperature, $C(T,N)$, is diagnostic for the underlying density of states, and describes how the states of the system are populated as the temperature increases. Two systems of the same size with similar $C(T,N)$ curves can thus be said to be similar to each other.   We recall that the heat capacity can be estimated from Monte Carlo simulations at a single temperature by computing the variance of the energy over many binary patterns generated by the simulation, $C(T,N) = \langle (\delta E)^2 \rangle / T^2$. 

The first interesting result is that when we choose a subnetwork of $N$ neurons at random out of the 40 cells from which we record, for $N=5,\, 10,\, 15,\,$ or $20$, we find that the fluctuations in $C(T,N)$ are very small.    This suggests that the individual networks, even when quite small, are typical of the ensemble of subnetworks that we can choose out of this small patch in the retina.

More ambitiously, we can try to construct an artificial ensemble of networks that reproduces the distribution of mean spike probabilities and the distribution of pairwise correlations that we see in the experiment.  Concretely, we assign to each neuron a mean spike probability chosen at random from the observed distribution of mean spike probabilities (Fig \ref{fx}c shows, implicitly, the cumulative distribution of $\langle\sigma_{\rm i}\rangle$), and we assign to each pair of cells a correlation chosen at random from the observed distribution of correlations, shown in Fig \ref{fx}e.
Note that not all combinations of means and correlations are possible for binary variables;  after each draw from the distribution of 
$\langle \sigma_{\rm i}\rangle$ and $C_{\rm ij}$, we check that all 
$2\times 2$ marginal distributions are consistent, and repeat the draw if needed. Once the whole synthetic covariance matrix is generated, we check (e.g. using the Kolmogorov--Smirnov test) that the distribution of its elements is consistent with the measured distribution.  Figure \ref{f2} shows $C(T,N)$ for networks of 20 neurons constructed in this way, and we see that, within error bars, the behavior of these randomly chosen systems resembles that of real 20 neuron groups in the retina. 


We emphasize that  ensemble we construct here is very different from the usual one in statistical mechanics.   In the theory of spin glasses \cite{mezard+al_87}, it is the interactions $J_{\rm ij}$ which are chosen at random.  Importantly, each pairwise interaction is chosen independently.  If we try to do this in our problem, we find a disaster---when we randomize the fields $h_{\rm i}$ and interactions $J_{\rm ij}$, we find that distribution of mean spike probabilities $\langle \sigma_{\rm i}\rangle$ changes dramatically, and as a result everything else about the network also changes (heat capacity, entropy, ... ).  In contrast, here  we keep the distribution of observables, i.e. the pairwise correlations and firing rates, fixed at their measured values, and moreover independent of $N$, as motivated by experiments that find no decay in the pairwise correlation in patches of $\sim 200$ neurons. 


 
 A striking feature of the spin glass problem is that when the $J_{\rm ij}$ are drawn independently and at random, the correlations among the spins have a rich, hierarchical structure  \cite{mezard+al_87}.  In our construction method  we draw the correlation matrix elements independently (subject to the consistency condition above), and therefore expect that this will induce a complicated correlation structure in the space of $J_{\rm ij}$ couplings.
 
   
We are building a spin glass model in which all pairs potentially interact, and all the pairwise correlations are drawn from the same distribution.  In this sense, we expect that we have some sort of mean--field model, but the correlations have a scale set by experiment, and hence cannot be reduced as $N$ becomes larger.  This family of models  clearly cannot  have a normal thermodynamic limit as $N$ becomes large. This is not a failing of the model, however:  the \emph{real} neural system has correlations that do not appreciably decay with distance in mesoscopic patches (presumably because of the extended connectivity of the neurons and correlations in the stimulus), and maximum entropy models can explore the consequences of these measured constraints. Recalling our results on the toy  model in Section \ref{sectoy}, we might expect our systems to be driven towards criticality as $N$ is increased; we explore this issue in the next section.


\section{Larger networks and criticality}

Figure \ref{f2} reveals an interesting behavior of heat capacity $C(T,N)$: as the size of the system increases, the peak of the heat capacity moves closer and closer to the operating point at $T=1$, in networks of size $N\leq 40$ constructed from data. 
Armed with the results at $N=20$ and an operational definition of a network ensemble, we decided to check if this behavior continues to larger $N$. We thus generated several
synthetic networks of 120 neurons by randomly choosing once more out of the distribution of $\langle\sigma_{\rm i}\rangle$ and $C_{\rm ij}$ observed experimentally.  The heat capacity $C(T,N=120)$ now has a dramatic peak at
$T^*=1.07\pm0.02$, very close to the operating point at $T=1$.
  
To be clear, we note that the shift in the peak of heat capacity with the system size $N$ is a direct consequence of pairwise couplings in the model, and hence this structure is driven by the measured correlation among neurons.
In independent models, a simple calculation shows that 
 \begin{equation}
C_{\rm ind}(T,N)=\sum_{{\rm i}=1}^N \left(h_{\rm i}/T\right)^2 \left[1-\tanh^2 (h_{\rm i}/T)\right],
\end{equation}
where $h_{\rm i}=\tanh^{-1} (\langle\sigma_{\rm i}\rangle)$;  for our dataset, the ensemble average of $C_{\rm ind}(T,N)$ for independent models peaks at about $T\sim 1.7$ regardless of $N$, and indeed, when normalized by $N$, the curves of $C_{\rm ind}(T,N)/N$ collapse onto a single curve.   In contrast, in pairwise models of the same data, the peak moves from $T\sim 1.7$ for $N=5$ to $T\sim 1.3$ for $N=40$, while the heat capacity per neuron, $C(T,N)/N$ increases by about 40\%.  This is yet another indication that, although the correlation between any two neurons is weak, neglecting these correlations gives a qualitatively wrong picture of the states accessible to the network as a whole.

In physical systems, a sharp peak in the heat capacity, becoming singular as $N$ becomes large, is associated with a critical point in the phase diagram.  The critical point is distinguished by the fact that at this point, the system is maximally sensitive to small changes in parameters.  It has been suggested that this sensitivity makes operation at a critical point attractive as a strategy for biological sensory and signaling systems  \cite{critical1,critical2,critical3}.  Behavior in the neighborhood of the critical point also is universal, so that systems with many different microscopic structures can generate, quantitatively, the same critical behavior.  It should be noted that there are different notions of criticality which have been applied to biological systems.  As far as we know this is the first evidence for criticality of a biological network in the thermodynamic sense.

From the thermodynamic point of view, the critical point marks a boundary between phases.  But from a statistical point of view, this point is at an extremum.  Specifically, in a large system we expect that almost all states which are accessible will have nearly equal probability; in information theory this is the idea of ``typicality'' (usually applied to long sequences), and in statistical mechanics this is the equivalence of the canonical and microcanonical ensembles.  The critical point is the place where the corrections to this expectation are largest.  More precisely, if we write the probability distribution in the Boltzmann form, as in Eq (\ref{ham}), the (negative) log of the probability of visiting  a state is the energy, and the heat capacity is proportional to the variance of the energy.  Thus, at the critical point, where the heat capacity has its maximal value, the variance of log probability is largest.

%
If  the brain is interested directly in how surprised it should be by the \emph{current} state of its inputs, then it might be important to maximize the dynamic range for instantaneously representing this (negative log) likelihood, that is, to have at its disposal a codebook with word frequencies whose range is wide enough to encompass the range of probabilities of sensory events.  
In contrast,  simple notions of efficient coding would require all symbols to be used with equal probability.  But since states of the visual world occur with wildly varying probabilities,  achieving this notion of efficiency would require block codes that are extended over time and are therefore slow.

\section{Entropy and multi-information}

Scaling of the temperature introduced in Section \ref{secens} is useful for another reason:  it enables us to estimate the entropy of $P(\left\{\sigma_{\rm i}\right\})$. The entropy in the context of coding  measures the capacity of the neurons to convey information about the visual world: the single-neuron biases and interactions effectively reduce the total number of likely binary patterns (or the codebook size) from $2^N$ to $2^{S(T=1,N)}$, and we would like to quantify this decrease.
We recall that   
\begin{equation}
S(N)=S(T=1,N) =\int_0^1 \frac{C(T,N)}{T}\,dT, \label{sint}
\end{equation}
where the heat capacity $C(T,N)$  can be estimated by Monte Carlo from the variance of the energy, $C(T,N)=\langle (\delta E)^2\rangle/T^2$, by drawing a large number of spin configurations at a fixed temperature $T$, computing their energy using the Hamiltonian of Eq (\ref{ham}) and taking the variance.  Then, the integral in Eq (\ref{sint}) is performed up to $T=1$, corresponding to the pairwise model of the real data. Estimating the entropy using the heat capacity integration in Eq (\ref{sint}) is crucial for $N>20$, because direct estimation from raw Monte Carlo samples becomes infeasible.

 \begin{figure} 
   \centering
   \includegraphics[width=3.3in]{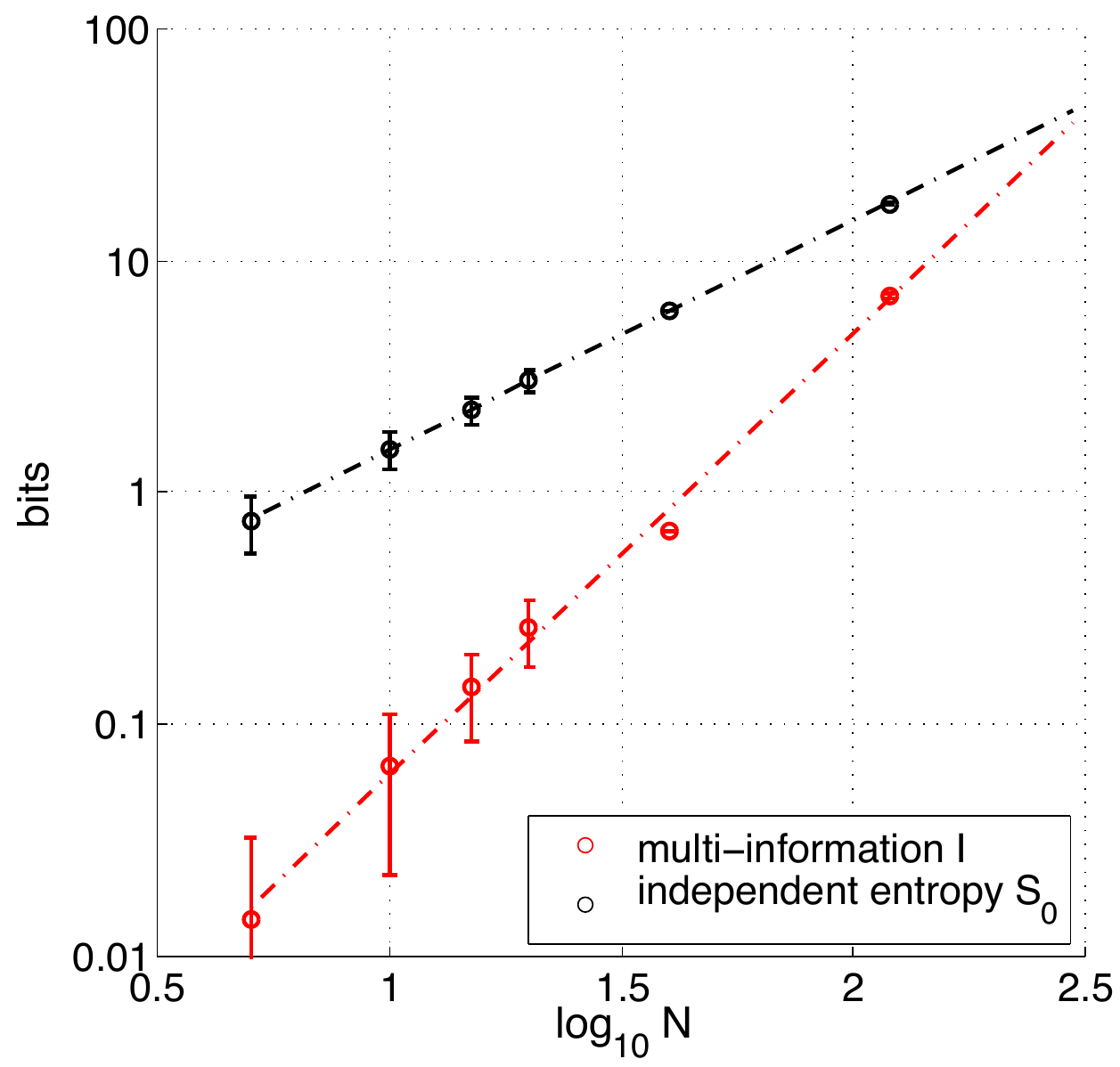}
   \caption{The scaling of multi-information $I$ and the independent entropy $S_0$ with the system size. Points at network size $N=5,10,15,20,40$ are computed from real data, with error bars denoting the spread across many subnetworks selected from $N=40$ (there is no error estimate for the single network at $N=40$). Dashed lines are scaling fits; independent entropy scales approximately linearly, and multi-approximation approximately quadratically, with system size $N$, and this scaling continues up to the three synthetic networks of size $N=120$.  } 
   \label{f6}
\end{figure}

If we integrate Eq (\ref{sint}) to find the entropy of the large synthetic networks, we find that   the independent entropy of the individual spins, $S_0 (120) = 17.8\pm 0.2\,{\rm bits}$,  has been reduced to $S(120) = 10.7\pm 0.2\,{\rm bits}$ by pairwise interactions.  As Fig \ref{f6} shows, even at $N=120$ the entropy deficit or multi--information $I(N) = S_0(N) - S(N)$ for the Ising models continues to grow in proportion to the number of pairs ($\sim N^2$), continuing the pattern found in smaller networks \cite{schneidman+al_06}. We also note that Ising models provide a lower bound on the total amount of multi-information in the real distribution, because they only capture the pairwise structure. Therefore, if higher-order couplings become important at larger $N$, the real $I(N)$ might be bigger that the one estimated in Fig \ref{f6}. The next generation of experiments \cite{dario}, probing the $N\sim 100 - 200$ regime, will provide a decisive test of the maximum entropy model predictions.

\section{Locally stable states}

In the Hopfield model, dynamics of the neural network corresponds to motion on an energy surface.  Simple learning rules can sculpt the energy surface to generate multiple local minima, or attractors, into which the system can settle.  These local minima can represent stored memories, or the solutions to various computational problems \cite{hopfield+tank_85,hopfield+tank_86}.  By analogy with spin glasses, we can think of these multiple attractors, or locally stable states, as resulting from the competition between positive and negative interactions.    In our maximum entropy models, we similarly find a range of $J_{\rm ij}$ values encompassing both signs, as shown in Figs \ref{fx}b and \ref{fx}f.  We would like to understand whether this structure leads to multiple attractors, and what this means for the nature of the neural code.

\begin{figure} 
   \centering
   \includegraphics[width=3.5in]{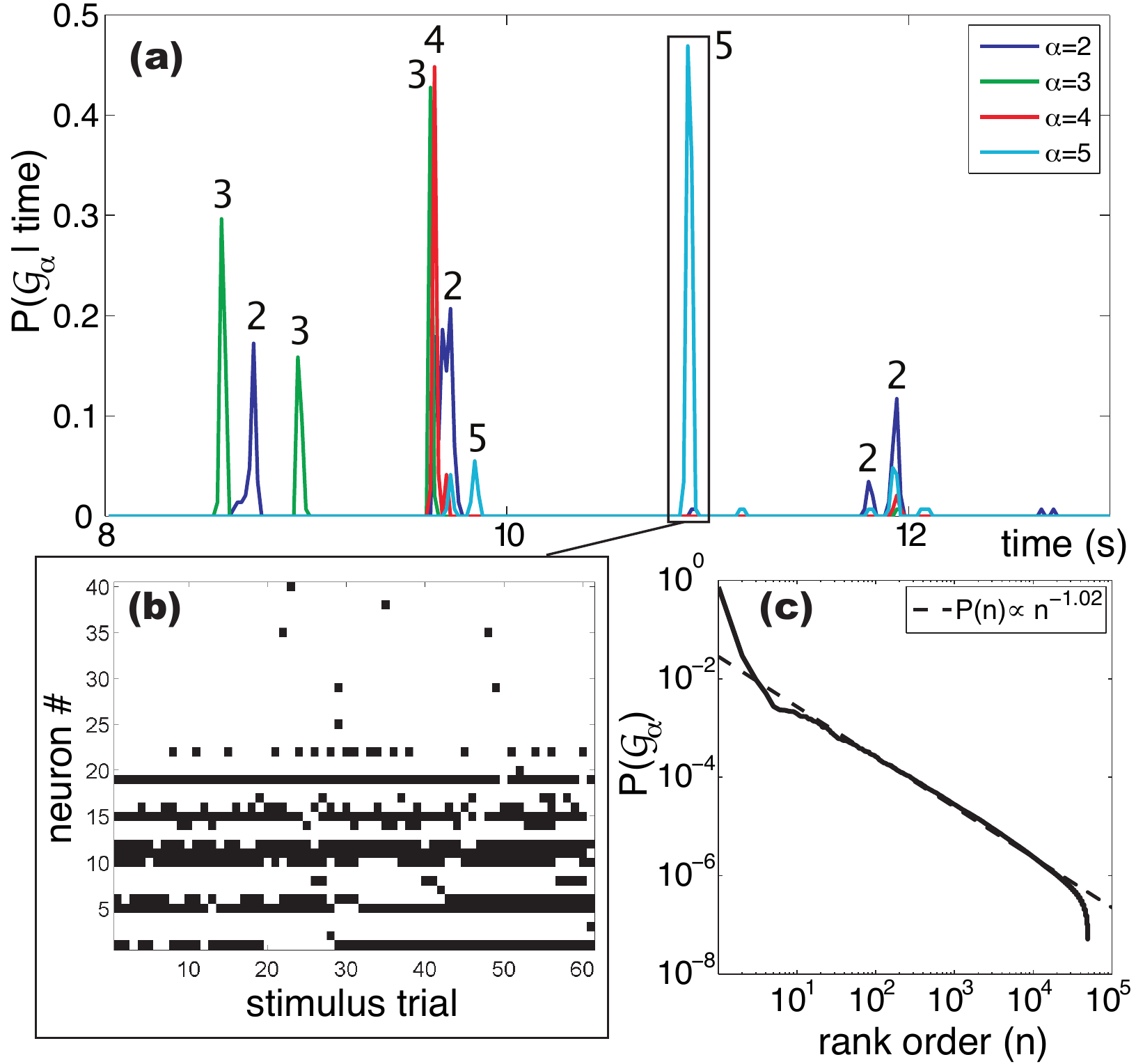}
   \caption{ (a) 
   Probability that the 40 neuron system is found in a configuration within the basin of attraction of each nontrivial locally stable state ${\cal G}_{\alpha}$, as a function of time during the stimulus movie; $P({\cal G}_{\rm \alpha} | t) = 0.4$ means that the retina is in that basin on $40\%$ of the 145 repetitions of the movie.
   (b)   All unique patterns  assigned to $\mathcal{G}_5$ at $t=10.88-10.92\e{s}$. (c) Zipf plot of the rank ordered frequencies with which the lowest lying $5\E{4}$ locally stable states are found in the simulated 120 neuron system.
   }
   \label{f4}
\end{figure}

Locally stable states are patterns of activity $\mathcal{G}_\alpha=\left\{\sigma_{\rm i}\right\}$, such that a flip of any single spin in $\mathcal{G}_\alpha$ increases the energy (or decreases the likelihood) of the new state.
At $N=40$ we find 4  such
local energy minima ($\mathcal{G}_2,\cdots ,\mathcal{G}_5$) in the observed sample that are stable against single spin flips, in addition to the silent state $\mathcal{G}_1$ ($\sigma_{\rm i}=-1$ for all $\rm i$). Using zero--temperature Monte Carlo, each configuration observed in the experimental data is assigned to 
its corresponding stable state: one starts with a binary pattern observed in the data, and flips the spins as long as each spin flip decreases the energy under the reconstructed Ising Hamiltonian $\mathcal{H}$. The local energy minimum thus found is matched to one of the $\mathcal{G}_\alpha$.   Although this assignment makes no reference to the visual stimulus,  the  collective states ${\cal G}_\alpha$ are reproducible across multiple presentations of the same movie (Fig \ref{f4}a), even when the microscopic state $\{\sigma_{\rm i}\}$ varies substantially (Fig \ref{f4}b). 

 In a simulated network of $N=120$ neurons, we find a much richer energy landscape. Looking in detail at  the distribution of   $J_{\rm ij}$, we find that it is approximately Gaussian  $\overline{J}=-0.016\pm 0.004$ and $\sigma_J=0.61\pm 0.04$, with
$53\%$ of triangles being frustrated ($46\%$ at $N=40$), indicating the possibility of many single-spin-flip stable states, as in spin glasses \cite{mezard+al_87}. 
For each $N=120$ network, we thus generated one long run collecting $2\E{7}$ independent samples. For each sample zero-temperature Monte Carlo is again used to determine the appropriate basin of attraction;  we tracked $5\E{4}$ lowest metastable states and kept detailed statistics for $10^3$ lowest.
We conclude that the Gibbs state now is a superposition of  thousands of $\mathcal{G}_\alpha$, with a nearly Zipf--like distribution (Fig \ref{f4}c). We checked that the probability of the Monte Carlo to generate a spin configuration that belongs to basin of attraction $\mathcal{G}_\alpha$ is to a good approximation proportional to $e^{-F_\alpha}$, where   the free energy $F_\alpha=\langle E\rangle_\alpha - S_\alpha$ is the difference between the average energy in the basin of attraction and its entropy, both estimated from the simulation data (i.e., from binary patterns that belong to the basin $\mathcal{G}_\alpha$). The average escape barrier size for the studied meta-stable states is $\sim 5 $, in natural units used in Eq (\ref{ham}).

The entropy of the distribution of metastable states, $P(\mathcal{G}_\alpha)$ is $3.4\pm 0.3 \,{\rm bits}$, about a third of the total entropy.  Thus, a substantial fraction of the network's capacity to convey visual information would be carried by the collective state, that is, by the identity of the basin of attraction, rather than by the detailed microscopic states. 

Based on these observations, we can formulate the following hypothesis: Trial-to-trial variability demonstrates that neurons must be to some extent noisy and  therefore each `microscopic' state or binary pattern $\left\{\sigma_{\rm i}\right\}$ cannot be a codeword with a separate meaning; instead, the space of binary patterns must be partitioned into domains or regions containing similar patterns with synonymous meanings. We propose that such domains are exactly the basins of attraction of locally stable states of $\mathcal{H}$. One desirable property of this choice is that the similarity metric for microscopic states is the energy function (or the likelihood) itself, which captures our intuition that the most frequently observed binary word among words that differ in a single letter is probably the `correctly spelled' variant, and the other variations are akin to spelling mistakes. The second desirable characteristic is that an associative and error correcting mechanism for parsing such words already exists, and is simply the Hopfield network.  We also find that on the synthetic network of $N=120$ neurons, observing $K\ll N$ neurons is on average approximately two times more informative about the collective state $\mathcal{G}_\alpha$ than about an unobserved neuron, i.e. $I(\vec{\sigma};\mathcal{G}_\alpha)/S(\mathcal{G}_\alpha)\sim 2 I(\vec{\sigma};\sigma_l)/S(\sigma_l)$, where $\vec{\sigma}$ is a group of $K\leq 10$ neurons that doesn't include $\sigma_l$, and the information is normalized by the corresponding total entropy. This indicates that the decoding  of  locally stable states  from a neural sub-population should be easier than decoding the microscopic pattern of activity.

To test this hypothesis, one would need to compare the information that basins of attraction provide about the stimulus $s$, $I(\mathcal{G_\alpha};s)$, with the information that the microscopic state provides about the stimulus, $I(\left\{\sigma_{\rm i}\right\};s)$. If these two quantities are similar in size, then basin-of-attraction identity $\mathcal{G}_\alpha$ is a good summary (or compression) of the microscopic state, and can be used to transmit information. Experiments could also be used  to refine the hypothesis by checking if, for example, the basin of attraction provides information about some identifiable invariant property of the stimulus (e.g., spatial shape), while additional information about other aspects of the stimulus (e.g., contrast) is encoded in the microscopic pattern within the  basin of attraction. Unfortunately, in the real networks of $N=40$ neurons we can just start to detect the emergence of the metastable states, and we are thus unable to perform the test. While at $N=120$ neurons the structure is much richer, these synthetic networks have no associated set of `experimentally observed' patters locked to the stimulus, so such a test, even simulated, can also not be performed. 
To conclude this section, we note that locally stable states can be defined without a reference to a particular model (the Ising model here), by simply finding patterns that are sampled well enough in the data and are more frequent than all of their single-spin-flip neighbors \cite{stephens}. Even if pairwise models were to break down at higher $N$, our suggestion might still provide a viable coding hypothesis.

\section{Discussion}

Ising models, with a spin glass structure of competing interactions, are the least structured models that can describe the observed mean spike probability and pairwise correlations in networks of real neurons.  Remarkably, these minimal models continue to provide a good description of the higher order correlations among retinal neurons up   to $N=40$.  Although correlations among pairs of cells are weak, the behavior of these large groups of cells is strongly collective \cite{collective}. 

In the Ising model, the bias for a particular neuron to fire (spin up) or remain silent (spin down) has two components, one intrinsic to the neuron and one from the interactions with other neurons.  At $N=40$, these different contributions are comparable, and so the pairwise models cannot simply be viewed as a small perturbative correction to the independent model.
Nevertheless, the sparsity of spikes is of crucial importance for models of neural behavior. In small networks at least, rare spiking means that higher order interactions don't have much chance to contribute even if they are present, and we hypothesize that this property of neural code could be important also as the brain tries (literally) to make sense of the incoming data.

Having found that the Ising model provides a good description of the real network, we are encouraged to take the model seriously as a statistical mechanics problem.    In particular, since the system has competing interactions that differ from neuron to neuron, we would like to understand, in the spirit of spin glass theory \cite{mezard+al_87}, whether the particular system that we observe is typical of some ensemble of networks.   We have been able to construct such an ensemble, and use this as a tool to predict the behavior of larger systems.  In the salamander retina, the ``correlated patch'' of neurons, within which the pairwise correlations are largely independent of the distance between cells, contains $N\sim 200$ cells, so we would like to understand what our framework predicts on this scale.    Remarkably, the larger networks that we construct seem to be operating close to a critical point in their phase diagram.

Criticality is perhaps the most dramatic signature of collective behavior.  The next generation of experiments should have access to the full population of $N\sim 200$ cells \cite{segev+al_04,gunning+al_05,dario}, and will test this prediction in detail.  As emphasized in recent work on natural images, one can find evidence of criticality even without constructing an explicit statistical model \cite{stephens}.  Further, it should be remembered that the pattern of correlations in the retina depends not just on the underlying connectivity or the structure of the visual inputs, but also on the adaptation state of the system.  While much has been learned about the way in which individual neurons adapt to changes in the input stimulus statistics, especially in the retina \cite{adapt1,adapt2}, much less is known about how these adaptation processes influence the behavior of whole networks.  If operation at a critical point is an essential feature of network function,   adaptation after a sudden change in input statistics should bring the system back to criticality, even if the exact pattern of spiking probability and correlations in the network is changed, and this is testable.   For a model of how adaptive dynamics can enforce critical behavior, see Ref \cite{magnasco+al_09}.



Our second prediction concerns the emergence and role of the locally stable states in the probability distribution of binary words, $P(\left\{\sigma_{\rm i}\right\})$: we have shown that at $N=40$ several locally stable states appear reproducibly from trial to trial despite substantial variability in the detailed binary patterns of neural activity.   Importantly, the analysis which identifies these states makes no reference to the visual stimulus, yet these states are tied to the stimulus in a way which suggests a role in the neural code.
Furthermore, by studying the synthetic networks we have demonstrated that the vocabulary of such states could vastly expand by the time $N$ reaches $\sim 100$, providing enough capacity and dynamic range even to dominate the encoding of incoming stimuli. Such a collective code is not inconsistent with the single-neuron results obtained to date; rather, one possible manifestation of collective coding is that, as more and more neurons are observed simultaneously, what was thought to be noise or uncorrelated fluctuations in a single neuron, is really a part of the collective state.

Both ideas---tuning to criticality and the use of stable states as robust codewords---have been motivated by experiments on a 40-neuron network where the beginnings of nontrivial collective behavior could be identified. We have outlined how these predictions can plausibly be tested in a next generation of experiments that will access networks of $N\sim 100$ or larger. If validated,  these and similar results  would provide a substantive link from real data to    to the large body of theoretical work on neural networks.

\begin{acknowledgments}
This work was supported in part by NIH  Grants R01 EY14196 and P50 GM071508, by the E. Matilda Zeigler Foundation, by NSF Grants IIS--0613435 and PHY--0650617, and by the Burroughs Wellcome Fund Program in Biological Dynamics.  WB also thanks his colleagues at the Rockefeller University and at the Universit\'a di Roma, La Sapienza, for their hospitality during portions of this work.
\end{acknowledgments}

%

%

\begin{thebibliography}{}
%
\bibitem{hopfield_82}
JJ Hopfield, Neural networks and physical systems with emergent collective computational abilities.
{\em Proc Nat'l Acad Sci  (USA)} {\bf 79,} 2554--8 (1982).
\bibitem{amit_89}
DJ Amit, 
{\em Modeling Brain Function:  The World of Attractor Neural Networks} (Cambridge University Press, Cambridge, 1989).
\bibitem{hertz_91}
J Hertz, A Krogh \& RG Palmer, {\em Introduction to the Theory of Neural Computation} (Addison Wesley, Redwood City, 1991).
\bibitem{schneidman+al_06}
E Schneidman, MJ Berry II, R Segev \& W Bialek,
Weak pairwise correlations imply strongly correlated network states in a neural population. 
{\em Nature} {\bf 440,} 1007-1012 (2006), arXiv:q-bio/0512013v1.
\bibitem{spikes}
F Rieke, D Warland, RR de Ruyter van Steveninck \& W Bialek, {\em Spikes: Exploring the Neural Code} (MIT Press, Cambridge, 1997).
\bibitem{jaynes_57}
ET Jaynes, Information theory and statistical mechanics. {\em Phys Rev} {\bf 106,} 620--630 (1957).
\bibitem{schneidman+al_03}
E Schneidman, S Still, MJ Berry II \& W Bialek,
Network information and connected correlations, {\em Phys  Rev Lett} {\bf 91,} 238701 (2003), arXiv:physics/0307072v1.
\bibitem{shlens+al_06}
J Shlens, GD Field, JL Gaulthier, MI Grivich, D Petrusca, A Sher, AM Litke \& EJ Chichilnisky,
The structure of multi-neuron firing patterns in primate retina.
{\em J Neurosci} {\bf 26,} 8254-66 (2006).
\bibitem{shlens+al_09}
J Shlens, GD Field, JL Gaulthier, M Greschner, A Sher, AM Litke \& EJ Chichilnisky,
The structure of large-scale synchronized firing in primate retina. 
{\em J Neurosci} {\bf 29,} 5022--31 (2009).
\bibitem{cosyne}
IE Ohiorhenuan \& JD Victor, Maximum entropy modeling of multi-neuron firing patterns in V1. Proceedings of 2007 Cosyne conference; 
http://cosyne.org.
\bibitem{yu}
S Yu, D Huang, W Singer \& D Nikolic, A small world of neuronal synchrony. {\em Cereb Cortex} {\bf 18,} 2891--2901 (2008).
\bibitem{tangetal}
A Tang, D Jackson, J Hobbs, W Chen, JL Smith, H PAtel, A Prieto, D Petruscam MI Grivich, A Sher, P Hottowy, W Dabrowski, AM Litke \& JM Beggs,  A maximum entropy model applied to spatial and temporal correlations from cortical networks {\em in vitro}. {\em J Neurosci} {\bf 28,} 505--518 (2008).
\bibitem{preprint}
G Tka\v{c}ik, E Schneidman, MJ Berry II \& W Bialek, Ising models for networks of real neurons. arXiv:q-bio/0611072 (2006).
\bibitem{Broderick+al_08}
T Broderick, M Dudik, G Tka\v{c}ik, RE Schapire \& W Bialek, Faster solutions of the inverse pairwise Ising problem.  arXiv:0712.2437 (2007).
\bibitem{monasson}
S Cocco, S Leibler \& R Monasson, Neuronal couplings between retinal ganglion cells inferred by efficient inverse statistical physics methods. {\em Proc Nat'l Acad Sci USA} {\bf 106,} 14058--62 (2009).
\bibitem{nirenberg}
S Nirenberg \& J Victor, Analyzing the activity of large populations of neurons: How tractable is the problem. {\em Curr Opin Neurobiol} {\bf 17,} 397--400 (2007).
\bibitem{roudi1}
Y Roudi, E Aurell \& JA Hertz, Statistical physics of pairwise probability models. {\em Front Comput Neurosci} {\bf 3,} 22 (2009).
\bibitem{roudi2}
Y Roudi, S Nirenberg \& PE Latham, Pairwise maximum entropy models for studying large biological systems: when they can work and when they can't. {\em PLoS Comput Biol} {\bf 5,} e1000380 (2009).
\bibitem{roudi3}
Y Roudi, J Trycha \& J Hertz, The ising model for neural data: model quality and approximate methods for extracting functional connectivity. {\em Phys Rev E} {\bf 79,} 051915 (2009).
\bibitem{bethge}
M Bethge \& P Berens, Near-maximum entropy models for binary neural representations of natural images. In {\em Advances in Neural Information Processing Systems  20,} J Platt et al eds, pp 97--104  (Cambridge, MA, MIT Press, 2008).
\bibitem{thierry}
M Mezard \& T Mora, Constraint satisfaction problems and neural networks: a statistical physics perspective. {\em J Physiol Paris} {\bf 103,} 107--113 (2009).
\bibitem{cessac}
B Cessac, H Rostro, JC Vasques \& T Vieville, How Gibbs distributions may naturally arise from synaptic adaptation mechanisms. {\em J Stat Phys} {\bf 136,} 565--602 (2009).
\bibitem{marre}
O Marre, SE Boustani, Y Fregnac \& A Destexhe, Prediction of spatio--temporal patterns of neural activity from pairwise correlations. {\em Phys Rev Lett} {\bf 102,} 138101 (2009).
\bibitem{ranganathan}
W Bialek \& R Ranganathan, Rediscovering the power of pairwise interactions. arXiv.org:0712.4397 (2007).
\bibitem{gtphd}
G Tka\v{c}ik, {\em Information Flow in Biological Networks} (Dissertation, Princeton University, 2007).
\bibitem{seno}
F Seno, A Trovato, JR Banavar \& A Maritan, Maximum entropy approach for deducing amino acid interactions in proteins. {\em Phys Rev Lett} {\bf 100,} 078102 (2008).
\bibitem{volkov}
I Volkov, JR Banavar, SP Hubbell \& A Maritan, Inferring species interactions in tropical forests. {\em Proc Nat'l Acad Sci USA} {\bf 106,} 13854--13859 (2009).
\bibitem{dhadialla}
PS Dhadialla, IE Ohiorhenuan, A Cohen \& S Strickland, Maximum--entropy network analysis reveals a role for tumor necrosis factor in peripheral nerve development and function. {\em Proc Nat'l Acad Sci USA} {\bf 106,} 12494--12499 (2009).
\bibitem{weigt}
M Weigt, RA White, H Szurmant, JA Hoch \& T Hwa, Identification of direct residue contacts in protein-protein interaction by message passing. {\em Proc Nat'l Acad Sci USA} {\bf 106,} 67--72 (2009).
\bibitem{segev+al_04}
R Segev, J Goodhouse, JL Puchalla \& MJ Berry II, Recording spikes from a large fraction of the ganglion cells in a retinal patch. {\em Nat Neurosci} {\bf 7,} 1155--62 (2004).
\bibitem{gunning+al_05}
D Gunning, C Adams, W Cunningham, K Mathieson, V O'Shea, KM Smith, EJ Chichilnisky, 30$\mu$m spacing 519-electrode arrays for in vitro retinal studies. {\em Nuc Inst Methods A} {\bf 546,} 148--153 (2005).
\bibitem{dario}
D Amodei, G Schwartz \& MJ Berry II, Correlations and the structure of the population code in a dense patch of the retina. {\em Proceedings MEA Meeting 2008}, pp. 197, Stett A ed., BIOPRO Baden-W\"urttemberg, 2008.
\bibitem{keat}
J Keat, P Reinagel, RC Reid \& M Meister, Predicting every spike: a model for the responses of visual neurons. {\em Neuron} {\bf 30,} 803--817 (2001).
\bibitem{fairhall_berry}
AL Fairhall, CA Burlingame, R Narasimhan, RA Harris, JL Puchalla \& MJ Berry II,  Selectivity for multiple stimulus features in retinal ganglion cells. {\em J Neurophysiol} {\bf 96,} 2724--2738 (2006).
\bibitem{pillow}
JW Pillow, J Shlens, L Paninski, A Sher, AM Litke, EJ Chichilnisky, EP Simoncelli, Spatio-temporal correlations and visual signalling in a complete neuronal population. {\em Nature} {\bf 454,} 995-9 (2008).
\bibitem{kolia}
KS Sadeghi, {\em Progress on Deciphering the Retinal Code.} (Dissertation, Princeton University, 2009).
\bibitem{perkel+bullock_68}
D Perkel \& T Bullock, Neural coding.  {\em Neurosci Res Program Bull} {\bf 6,} 221--343 (1968). 
\bibitem{ginzburg+sompolinsky_94}
I Ginzburg \& H Sompolinsky, Theory of correlations in stochastic neural networks. {\em Phys Rev E } {\bf 50,} 3171--3191 (1994).
\bibitem{latham+nirenberg} 
S Nirenberg, SM Carcieri, AL Jacobs \& PE Latham, Retinal ganglion cells act largely as independent encoders. {\em Nature} {\bf 411,} 698--701 (2001).
\bibitem{hopfield_08}
JJ Hopfield, Searching for memories, sudoku, implicit check bits, and the iterative use of not--always--correct rapid neural computation.  {\em Neural Comp} {\bf 20,} 1119--1164 (2008); arXiv:q--bio/0609006.

\bibitem{field}
D Field, Relations between the statistics of natural images and the response properties of cortical cells. {\em J Opt Soc A} {\bf 4,} 2379--94 (1987).
\bibitem{ruderman}
DL Ruderman \& W Bialek, Statistics of natural images: scaling in the woods. {\em Phys Rev Lett} {\bf 73,} 814--817 (1994).
\bibitem{dong+atick} 
DW Dong \& JJ Atick, Statistics of natural time-varying images. {\em Network} {\bf 6,} 345--358 (1995).
\bibitem{simoncelli+schwartz}
O Schwartz \& EP Simoncelli, Natural signal statistics and sensory gain control. {\em Nat Neurosci} {\bf 4,} 819--25 (2001).
\bibitem{stephens}
GJ Stephens, T Mora, G Tka\v{c}ik \& W Bialek, Thermodynamics of natural images. arXiv.org:0806.2694 (2008).
\bibitem{kadanoff}
LP Kadanoff, {\em Statistical Physics: Statics, Dynamics and Renormalization} (World Scientific, Singapore, 2000).
\bibitem{puchalla+al_05}
JL Puchalla, E Schneidman, RA Harris \& MJ Berry II,
Redundancy in the population code of the retina.
{\em Neuron} {\bf 46,} 493--504 (2005).
\bibitem{segev_physiol}
R Segev,  J Puchalla \& MJ Berry II,
Functional organization of retinal ganglion cells in the salamander retina. {\em J Neurophysiol} {\bf 95,} 2277--92 (2006).

\bibitem{Hinton}
GE Hinton \& TJ Sejnowski,
Learning and relearning in Boltzmann machines.  In {\em Parallel Distributed Processing: Explorations in the Microstructure of Cognition, Vol. 1,} ed. DE Rumelhart, JL McClelland \& the PDP Research Group, pp. 282--317 (MIT Press, Cambridge, 1986).
\bibitem{dudik}
M Dudik, SJ Phillips \& RE Schapire, Performance guarantees for regularized maximum entropy density estimation. {\em Proceedings 17th Annual conference on learning theory} (2004).

\bibitem{cd}
GE Hinton, Training products of experts by minimizing contrastive divergence. {\em Neural Comput} {\bf 14,} 1771--1800 (2002).

\bibitem{lin_91}
J Lin, Divergence measures based on the Shannon entropy. {\em IEEE Trans Inf Theory} {\bf 37,} 145--151 (1991). 
%
\bibitem{sessak}
V Sessak \& R Monasson, Small-correlation expansions for the inverse Ising problem. {\em J Phys A} {\bf 42,} 055001 (2009).
%
\bibitem{azhar_thesis}
F Azhar, {\em An Information Theoretic Study of Neural Populations} (Dissertation, University of California at  Santa Barbara, 2008).

\bibitem{mezard+al_87}
M Mezard, G Parisi \& MA Virasoro, 
{\em Spin Glass Theory and Beyond} (World Scientific, Singapore, 1987).

\bibitem{critical1}
TAJ Duke \& D Bray, Heightened sensitivity of a lattice of a membrane receptors. {\em Proc Nat'l Acad Sci (USA)} {\bf 96,} 10104--8 (1999).
\bibitem{critical2}
VM Eguiluz, M Ospeck, Y Choe, AJ Hudspeth \& MO Magnasco,
Essential nonlinearities in hearing.
{\em Phys Rev Lett}  {\bf 84,} 5232--5 (2000).
\bibitem{critical3}
S Camalet, T Duke, F J\"ulicher \& J Prost, Auditory sensitivity provided by self-tuned critical oscillators of hair cells. {\em Proc Nat'l Acad Sci (USA)} {\bf 97,} 3183-3188 (2000).
\bibitem{hopfield+tank_85}
JJ Hopfield \& DW Tank, ``Neural'' computation of decisions in optimization problems.  {\em Biol Cybern} {\bf 52,} 141--152 (1985).
\bibitem{hopfield+tank_86}
JJ Hopfield \& DW Tank, Computing with neural circuits: a model. {\em Science} {\bf 233,} 625--633 (1986).
\bibitem{collective}
For results on model systems cf. SM Bohte, H Spekreijse \& PR Roelfsema, The effects of pairwise and higher-order correlations on the firing rate of a postsynaptic neuron. {\em Neural Comput} {\bf 12,} 153--179 (2000).
\bibitem{adapt1}
S Smirnakis, MJ Berry II, D Warland, W Bialek \& M Meister, Adaptation of retinal processing to image contrast and spatial scale. 
 {\em  Nature} {\bf 386,} 69--73 (1997).
 \bibitem{adapt2}
 T Hosoya, SA Baccus \& M  Meister,  Dynamic predictive coding by the retina. {\em Nature} {\bf 436,} 71--77 (2005).


\bibitem{magnasco+al_09}
MO Magnasco, O Piro \& GA Cecchi, Self--tuned critical anti--Hebbian networks.  {\em Phys Rev Lett} {\bf 102,} 258102  (2009); 	arXiv:0808.3996v1.





\end{thebibliography}
 \end{document}